\documentclass[%
 reprint,
 amsmath,amssymb,
 aps,
]{revtex4-2}

\usepackage{graphicx}
\usepackage{dcolumn,color,multirow}
\usepackage{bm}

\begin{document}


\title{Full-scatter vector field analysis of an overmoded and periodically-loaded \\
cylindrical structure for the efficient transportation of THz radiation}

\author{Adham~Naji$^1$}\email{anaji@scu.edu}
 \altaffiliation[]{}
\author{Pawan~Kumar~Gupta$^1$}
\author{Gennady~Stupakov$^2$}
 
\affiliation{%
 $^1$Santa Clara University, Santa Clara, California, 95053\\
 $^2$SLAC National Accelerator Laboratory, Stanford, California, 94025
}%

\date{\today}

\begin{abstract}
Highly overmoded and periodically loaded structures, such as the iris-line waveguide, offer an attractive solution for the efficient transportation of diffraction-prone THz pulses over long distances (hundreds of meters). This paper presents the full-scatter field theory that allows us to analytically derive all the spectral (modal) coefficients on the discontuities of the iris line. The spectral analysis uses vector fields, superseding scalar field descriptions, to account for diffraction loss as well as polarization effects and ohmic loss on practical conductive surfaces. An advanced application of Lorentz's reciprocity theory, using a generalized guided-field configuration, is developed to reduce complexity of the mode-matching problem over nonuniform sections. The used technique is quite general and applies to a wide class of structures, as it only assumes a paraxial incidence (i.e.~a parabolic wave equation) along the axis of the structure. It removes the traditional assumption of very thin screens, allowing for the study of thicker screens in the high-frequency limit, while formulating the problem efficiently by scattering matrices whose coefficients are found analytically. The theory agrees with and expands previously established techniques, including Vainstein's asymptotic limit and the forward-scatter approximation.  The used formulation also facilitates accurate visualization of the transient regime at the entrance of the structure and how it evolves to reach steady state.
\end{abstract}

\maketitle


\section{Introduction} \label{Sec: intro}

The long-distance transportation of THz radiation, such as that generated by THz wigglers for pump-probe experiments in accelerator-based light-source facilities, is a challenging problem that has been recently under investigation \cite{DESY,NajiTHz1,NajiTHz2,Zhang}. This stems from working in a regime where the wavelength, $\lambda$, at THz frequencies is relatively so small as to render traditional (single-moded) wave-guiding solutions from the microwave-theory literature rather impractical, but not quite small enough as to eliminate its vulnerability to Fresnel diffraction. The overmoded iris-line waveguide is identified as one of the promising solutions that can be used to address such a challenge \cite{DESY,NajiTHz1,NajiTHz2}. The geometrically simple construct of this waveguide leverages the diffraction of the THz fields, over a series of irises, to transport dipolar (almost linearly polarized) radiation, while exhibiting low propagation losses and preserving linear polarization and an almost-Gaussain intensity profile across its output aperture \cite{DESY,NajiTHz1}.  

Practical overmoded structures that can guide high-frequency waves tend to be challenging to model, however, as they typically belong to a regime of analysis that lies between the limit of single-moded guided propagation and the limit of heavily diffractive propagation. The first limit assumes a relatively low frequency, as in traditional microwave waveguides or corrugated structures that offer phase-velocity control (slow-wave ring-loaded structures in radiofrequency charged-particle accelerators is an example). In this limit, one dominant mode is typically assumed, at least in some regions of the structure, and the analysis usually uses vector fields to account for polarization and conductive surface effects \cite{jackson,SchwingerLevine,schwingerWaveguides,Collin2,zangwill,Borgnis,pozar,Mahmoud,Vainstein1,slaterbook}. The second limit, on the other hand, tends to be concerned with thin edges and geometries, often with absorptive surface material, at higher frequencies. It typically uses scalar fields to describe diffraction effects, which can be well approximated asymptotically in some scenarios \cite{jackson,SchwingerLevine,schwingerWaveguides,Collin2,zangwill,Borgnis,pozar,Mahmoud,Vainstein1}. As boundary-value problems, we often analyze waveguiding structures spectrally (using a linear combination of the eigenmodes of the Helmholtz operator) or using integral equation representations (e.g.~Green functions, factorization, or Wiener-Hopf techniques); examples of which were pioneered in the seminal work by J.~Schwinger, L.A.~Vainetein, and others \cite{jackson,SchwingerLevine,schwingerWaveguides,Collin2,zangwill,Borgnis,pozar,Mahmoud,Vainstein1, Vainstein2, Dudley}. 
\begin{figure}
\centering
\includegraphics[width=0.84\columnwidth]{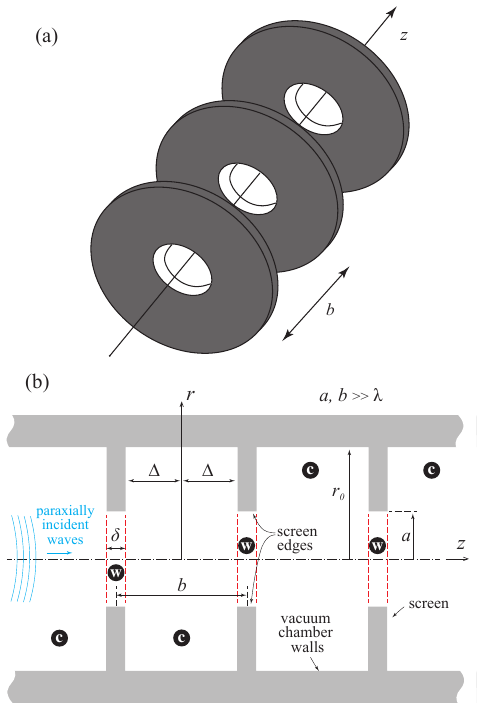}
\caption{A sketch showing the geometry of the iris line and the definition of key dimensions: (a) a three-dimensional sketch of the structure, without the enclosing vacuum chamber, (b) a cross-section, including the vacuum chamber. We seperate the problem domain into two types of regions (subdomains): we call the regions with narrower radius ($r\leq a$) ``waveguide" regions (indicated by a circle with the letter w in the figure), whereas regions with larger radius ($r\leq r_0$) are ``cavity" regions (indicated by a circle with the letter c in the figure). The red dashed lines in the figure represent the planes for the ``step-in" discontinuities (when traveling from a cavity to a waveguide region) and the ``step-out" discontinuities (when traveling from a waveguide to a cavity region).  }
\label{fig: geometry}
\end{figure}

When faced with a structure whose geometric features are oversized compared to the wavelength, $\lambda$, implying multi-mode (overmoded) operation, and made out of conductive material of finite thickness, $\delta\geq \lambda$, the required vector field analysis becomes more difficult. Standard numerical techniques, such as finite-element methods, may also become impractical due to the relatively large size of the structure compared to $\lambda$, forcing us to seek an analytical framework, not only to obtain a better insight into the wave physics of the structure, but also to speed up field computations.  Such a situation has been encountered during a study at the SLAC National Accelerator Laboratory, Stanford University, to transport a short THz pulse in the 3--15~THz range over 150--300~m for use in pump-probe experiments (with a similar consideration at the European X-ray Free-Electron Laser, European XFEL, in Germany) \cite{DESY,NajiTHz1,NajiTHz2,Zhang, DESY3,DESY4}. The highly overmoded iris line shown in Figure~\ref{fig: geometry} is one of the candidate solutions to transport this THz radiation without incurring too much loss to diffraction \cite{DESY,NajiTHz1,NajiTHz2}. When the screens are relatively thin, diffraction loss tends to be the dominant power loss mechanism during transportation \cite{DESY,NajiTHz1,NajiTHz2}. Note that, for such oversized structures and short THz pulse applications, any part of the wave energy that is diffracted or scattered from the axial transportation direction (along $z$ in Figure~\ref{fig: geometry}) into the shadow regions between the screens is considered lost. This allows one to model the conductive closed-chamber structure also as an open structure ($r_0\rightarrow \infty$), or a closed one with an absorptive chamber and screen side-walls \cite{NajiTHz1,NajiTHz2}.  Additionally, when analyzing the propagation of the short THz pulse in such a system (e.g.~for typical pump-probe experiments), note that we are chiefly concerned with the passage of the main (first) pulse through the waveguide, not with any delayed minor echoes that may potentially follow the pulse due to multiple reflections inside the waveguide geometry (if any). Indeed, as will be clear from the discussion in this paper and from references \cite{NajiTHz1,NajiTHz2}, even if these echoes existed, they are expected to be considerably attenuated and delayed in time compared to the main THz pulse (delayed at least by $\Delta t\geq 2cb\gg\tau$, where $\tau$ is the pulse width, $c$ is the speed of light, and $b$ is the period of the iris line).

Theoretical techniques, such as the eigen analysis based on mode matching \cite{NajiTHz1} and the forward-scatter field analysis \cite{NajiTHz2}, were recently developed to take into account realistic screens with finite values of $\delta$ and to investigate the effects of diffraction loss as well as any additional ohmic loss. The former method solves an inifinitely long structure for its steady-state eigenmodes, while the latter formulates each periodic section (cell) as a scattering problem, allowing flexibility in the analysis/design of finite length systems, while also boosting computational efficiency. The latter method, however, makes the assumption that the fields in each cell are scattered in the forward ($+z$) direction only, with no backward-traveling reflections \cite{NajiTHz2}. Although this assumption is reasonable at the high-frequency limit, its main advantage is in reducing the complexity of the resulting equations so that we can readily extract the field coefficients using mode orthogonality properties. 

A more rigorous analysis, however, will include a full-scatter formulation, where reflections are included and the coefficients of the spectral (modal) expansions of the fields are analytically produced. Conducting such an analysis will not only provide the most accurate results, but will also validate our usage of the forward-scattering approximation \cite{NajiTHz2}. To tackle the complexity of the full-scatter problem, we develop a special field formulation with the help of Lorentz's reciprocity theorem, a theorem by Schelkunoff for source equivalence from scattered field theory, and the field uniqueness theorem \cite{schelkunoff,jackson,Collin2,pozar,schwinger, Stratton}. To the best of our knowledge, the complete analytical results derived in this paper for the full-scatter spectral (modal) analysis, which are applicable to a general class of overmoded and periodically loaded cylindrical guiding structures, have not been reported before. 

The paper is organized as follows. In Section~\ref{Sec: analysis} we present the theoretical formulation of the problem in terms of full-scatter fields and obtain the analytical solutions for the field coefficients. Section~\ref{sec: implementations} then discusses the implementation of the theory using numerical examples, discusses the observed loss trends as a function of thickness, $\delta$, and demonstrates the agreement of the theory with previously established approximations and limits. The paper is concludes in Section~\ref{sec: conclusions}. Throughout the paper, we assume a time-harmonic dependence of the form $e^{-i\omega t}$, where $\omega$ is the angular frequency, $t$ is time, and $i\equiv\sqrt{-1}$.

\section{Analytical formulation} \label{Sec: analysis}

\subsection{Preliminary definitions and outline}

The total field expressions, including both the transverse electric (TE) and transverse magnetic (TM) mode expansions, in the waveguide regions are given by
\begin{align}
   E_{r,\text{TE}}^{\pm,w}&=\sum^{\infty}_{n=1} \frac{A^{\pm}_{n} a}{\nu'_{1n} r} J_{1}(\nu'_{1n}\frac{ r}{a})\cos\phi \ \zeta_\text{TE} \label{ArhoTE}
     \end{align}
\begin{align}
   E_{\phi, \text{TE}}^{\pm,w}&= -\sum^{\infty}_{n=1} A^{\pm}_{n} J'_{1}(\nu'_{1n}\frac{ r}{a})\sin\phi \ \zeta_\text{TE}  \label{AphiTE}\\
 E_{z,\text{TE}}^{\pm,w}&=0  \label{AzTE}\\
   H_{r,\text{TE}}^{\pm,w}&=\mp\sum^{\infty}_{n=1}  \frac{A^{\pm}_{n}}{Z_{0}}\left(\frac{\nu'^{2}_{1n}}{2k^{2}a^{2}}-1 \right)J'_{1}(\nu'_{1n}\frac{ r}{a})\sin\phi \ \zeta_\text{TE} \label{HrTE}\\
  H_{\phi, \text{TE}}^{\pm,w}&=\mp \sum^{\infty}_{n=1}  \frac{A^{\pm}_{n} a \left(\frac{\nu'^{2}_{1n}}{2k^{2}a^{2}}-1 \right)}{Z_{0} r\nu'_{1n}}J_{1}(\nu'_{1n}\frac{ r}{a})\cos\phi \ \zeta_\text{TE} \label{HphiTE}\\
   H_{z,\text{TE}}^{\pm,w}&=-\sum^{\infty}_{n=1} \frac{iA^{\pm}_{n}\nu'_{1n}}{Z_{0}k a}J_{1}(\nu'_{1n}\frac{ r}{a})\sin\phi \ \zeta_\text{TE} \label{HzTE}\\
       E_{r,\text{TM},n}^{\pm,w}&=-\sum^{\infty}_{n=1}B^{\pm}_{n}J'_{1}\left( \nu_{1n}\frac{ r}{a}\right)\cos\phi \ \zeta_\text{TM} \label{ArhoTM}\\
   E_{\phi,\text{TM},n}^{\pm,w}&=\sum^{\infty}_{n=1}\frac{B^{\pm}_{n} a}{\nu_{1n} r}J_{1}\left( \nu_{1n}\frac{ r}{a}\right)\sin\phi \ \zeta_\text{TM}  \label{AphiTM}\\
E_{z,\text{TM},n}^{\pm,w}&=\pm\sum^{\infty}_{n=1} \frac{iB^{\pm}_{n}\nu_{1n}}{ak}J_{1}\left( \nu_{1n}\frac{ r}{a} \right)\cos\phi \ \zeta_\text{TM}  \label{AzTM}\\
    H_{r,\text{TM},n}^{\pm,w}=&\mp\sum^{\infty}_{n=1} \frac{B^{\pm}_{n} a \left(\frac{\nu^{2}_{1n}}{2a^{2}k^{2}}+1 \right)}{Z_{0}\nu_{1n} r} J_{1}(\nu_{1n}\frac{ r}{a})\sin\phi \ \zeta_\text{TM}  \label{HrTM}\\
    H_{\phi,\text{TM},n}^{\pm,w}&=\mp\sum^{\infty}_{n=1} \frac{B^{\pm}_{n}\left(1+\frac{\nu^{2}_{1n}}{2a^{2}k^{2}}\right)}{Z_{0}}J'_{1}(\nu_{1n}\frac{ r}{a})\cos\phi \ \zeta_\text{TM} \label{HphiTM}\\
    H_{z,\text{TM},n}^{\pm,w}&=0 \label{HzTM}
\end{align}
where $\zeta_\text{TE}=e^{\pm(-iz\frac{\nu'^{2}_{1n}}{2ka^{2}}+ikz)}$, $\zeta_\text{TM}=e^{\pm(-iz\frac{\nu^{2}_{1n}}{2ka^{2}}+ikz)}$, $\phi$ is the azimuthal angle in cylindrical coordinates, $Z_0=120\pi$~$\Omega$ is the impedance of free space, $J_1$ denotes Bessel's function of the first kind and first order, $J'_1$ is its derivative with respect to its argument, $\nu_{1n}$ is the $n$th zero of $J_1$, $\nu'_{1n}$ is the $n$th zero of $J'_1$, and the structure dimensions ($a,b,r_0,\delta$) are defined in Figure~\ref{fig: geometry}. The superscripts ``w",``c" and ``in" are used to denote a waveguide-region field, a cavity-region field and an incident-mode field, respectively. See Figure~\ref{fig: geometry} for the definition of the waveguide/cavity regions and  step-in/step-out discontinuities. We use $A^\pm_n$ or $B^\pm_n$ to denote the coefficients of the $n$th TE or TM mode, respectively, with the $\pm$ sign in the superscript to indicate forward ($+z$) or backward ($-z$) travel direction. Note that (\ref{ArhoTE})--(\ref{HzTM}), which combine the forward-traveling and backward-traveling fields, are derived from first principles using the method outlined in \cite{NajiTHz2} for the forward-traveling fields. To include the backward-traveling reflections during derivation, the wavenumber $k$ is replaced by $-k$. The backward-traveling field expressions will have the same magnitude coefficients as those for the forward-traveling fields, except for the transverse magnetic ($H_r,H_\phi$) and longitudinal electric ($E_z$) components, which will have the same magnitude but an opposite sign compared to the forward-travel case. 

The useful symmetric properties of the field equations (\ref{ArhoTE})--(\ref{HzTM}) can be made clear during derivations by using a notation where we isolate the terms in front of each coefficient ($A^\pm_n$ or $B^\pm_n$) as normalized expression for each mode. We denote these normalized expressions by a hat overscript. For example, in (\ref{ArhoTE}) we can define $\hat{E}_{r,\text{TE},n}^{+,w}\equiv\frac{a}{\nu'_{1n} r} J_{1}(\nu'_{1n}\frac{ r}{a})\cos\phi \ \zeta_\text{TE}$, and therefore the $n$th mode can be written as $E_{r,\text{TE}}^{\pm,w}=A^{\pm}_{n}\hat{E}_{r,\text{TE},n}^{+,w}$, and so forth. Using this notation, (\ref{ArhoTE})--(\ref{HzTM}) can be rewritten in a reduced form at $z=0$ as
\begin{align}
  E_{r,\text{TE},n}^{\pm,w}=A^{\pm}_{n} \hat{E}_{r,\text{TE},n}^{+,w},\  &E_{r,\text{TM},n}^{\pm,w}=B^{\pm}_{n}\hat{E}_{r,\text{TM},n}^{+,w}\label{Er_reduced}\\
   E_{\phi, \text{TE},n}^{\pm,w}= A^{\pm}_{n} \hat{E}_{\phi, \text{TE},n}^{+,w},\ & E_{\phi,\text{TM},n}^{\pm,w}= B^{\pm}_{n}\hat{E}_{\phi,\text{TM},n}^{+,w}\label{Ephi_reduced}\\
   E_{z, \text{TE},n}^{\pm,w}=0,\  & E_{z,\text{TM},n}^{\pm,w}=\pm B^{\pm}_{n}E_{z,\text{TM},n}^{+,w}\label{Ez_reduced} \\
   H_{r,\text{TE},n}^{\pm,w}=\pm A^{\pm}_{n}\hat{H}_{r,\text{TE},n}^{+,w}, & H_{r,\text{TM},n}^{\pm,w}=\pm B^{\pm}_{n}H_{r,\text{TM},n}^{+,w} \label{Hr_reduced}\\
   H_{\phi, \text{TE},n}^{\pm,w}{=}\pm A^{\pm}_{n}\hat{H}_{\phi, \text{TE},n}^{+,w}, & H_{\phi,\text{TM},n}^{\pm,w}{=}\pm B^{\pm}_{n}H_{\phi,\text{TM},n}^{+,w} \label{Hphi_reduced}\\
   H_{z,\text{TE},n}^{\pm,w}=A^{\pm}_{n}\hat{H}_{z,\text{TE},n}^{+,w},\  & H_{z,\text{TM},n}^{\pm,w}=0 \label{Hz_recued}
\end{align}

Note that the same equations are obtained for the cavity regions, if we replace the radius $a\rightarrow r_0$ and the superscript $w\rightarrow c$.

Conducting a spectral (modal) analysis of the structure typically relies on expanding the locally scattered (or diffracted) field analytically in each region of the structure and matching such expansions at the discontinuity boundaries. This has been performed previously \cite{NajiTHz2}, where a single mode that is paraxially incident from the left side ($z<0$) upon a discontinuity (step-in or step-out) positioned at $z=0$ is matched to the modal expansion of field on the right side ($z>0$). This method assumes no backward-traveling reflections (forward-scatter is considered dominant). The summation of one side of the equation is then reduced to one term with the help of mode orthogonality properties \cite{NajiTHz2}, allowing us to find the analytical expressions for the coefficients $(A^+_n, B^+_n)$. The problem, however, becomes much more difficult when we try to  derive the $(A^\pm_n, B^\pm_n)$ coefficients in a full-scattered field analysis. The different mode summations with different radii in the Bessel function arguments will then exist on each side of the equation, rending the usual orthogonality properties insufficient by themselves to reduce the problem.  Traditional treatments seen in the literature often match modal expansions in uniform waveguides that have the same cross section on the left and right sides of a source region (singularity), or match them across discontinuous cross sections that assume only one (dominant) mode on one side,  \cite{jackson,Collin2,pozar,slaterbook}. In these cases, the analysis is easier and the modal sums are typically reduced using orthogonality integrals or a simple formulation of Lorentz's reciprocity theorem (discussed below). However, if a discontinuity has a cross-section change and many permissible modes are being scattered, literature treatments typically leave the result in terms of a sum of modes, without analytically finding the coefficients of these modes in explicit form. Alternative formulations, using variational, integral, or numerical techniques (e.g.~\cite{schwingerWaveguides, Collin2, jackson}) are sometimes used in place of spectral analysis for such discontinuities.  

In the following Subsection, however, we show that one way of circumventing this difficulty is: (1) by carefully choosing an appropriate setup for the second field term $(\mathbf{E}_2,\mathbf{H}_2)$ in Lorentz's reciprocity theorem, even when reflections are present and the cross-section is not uniform; (2) by utilizing a theorem by Schelkunoff for scattered field source equivalence, which allows us to decouple the different radii on each side of the discontinuity when applying orthonognality integrals; and (3) by utilizing uniqueness theorem for the field in a region with known source terms at the boundaries.  

\begin{figure}
\centering
\includegraphics[width=1\columnwidth]{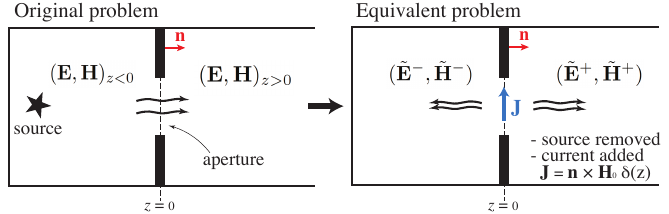}
\caption{A sketch demonstrating Schelkunoff's theorem. The original problem has a source and a field that fills the structure, passing through the aperture (opened in a conductive diaphragm, as indicated by the thick black lines). This original problem is replaced by an equivalent problem, whereby the original source is removed and an equivalent current source $\mathbf{J}=\hat{n}\times\mathbf{H}_{0}\delta(z)$ is placed in the aperture's plane. The equivalent current density $\mathbf{J}$ will produce scattered fields $(\tilde{\mathbf{E}}^-,\tilde{\mathbf{H}}^-)$ to the left, and $(\tilde{\mathbf{E}}^+,\tilde{\mathbf{H}}^+)$ to the right.}
\label{Schelk}
\end{figure}

\subsection{A discussion of field uniqueness and other useful theorems by Schelkunoff and Lorentz}

Schelkunoff's theorem \cite{schelkunoff,Collin2}, which originates from one of Schelkunoff's field scattering theorems (see Theorem no.~3 for electromagnetic waves in \cite{schelkunoff}), allows us to model a discontinuity by an equivalent electric current density source, $\mathbf{J}$, a magnetic current density source, $\mathbf{M}$, or both. With reference to Figure~\ref{Schelk}, we can obtain the fields at the LHS and RHS of the discontinuity, due to the original source by
\begin{eqnarray}
    (\mathbf{E},\mathbf{H})_{z<0} &=& (\mathbf{E}_0,\mathbf{H}_0) + (\tilde{\mathbf{E}}^-,\tilde{\mathbf{H}}^-) \label{Shelk-left field}\\
    (\mathbf{E},\mathbf{H})_{z>0} &=& (\tilde{\mathbf{E}}^+,\tilde{\mathbf{H}}^+),\label{Shelk-right field}
\end{eqnarray}
where the $(\mathbf{E}_0,\mathbf{H}_0)$ is the field found when the $z>0$ region is filled with a perfect electric conductor (PEC), and the scattered fields $(\tilde{\mathbf{E}}^\pm,\tilde{\mathbf{H}}^\pm)$, denoted by a tilde symbol in the overscript, are the forward/backward-traveling fields generated by an equivalent current source at z=0 derived from the fields when the PEC is present; namely,
\begin{equation}
\mathbf{J}(x,y,z)=\hat{n}\times\mathbf{H}_0(x,y,0)\ \delta(z),    \label{Schek J source}
\end{equation}
 where $\delta(z)$ is the Dirac delta function in $z$. Thus, the need to match the field expansions in the $z>0$ and $z<0$ regions directly is removed by applying a singularity (current sources) at the interface. 
 
 In using this formalism, we have implicitly utilized uniqueness theorem \cite{Stratton,jackson}, which allows us to get a unique solution to the coupled Helmholtz wave equation or Maxwell equations for the fields in a volume, if we have a complete description of either $\mathbf{J}$, $\mathbf{M}$, or both source terms on the entirety of the enclosing surface walls. The choice of $\mathbf{J}$ is therefore sufficient and convenient in the presence of the PEC ($z>0$) and due to the conductive wall of the aperture of the original problem in Figure~\ref{Schelk}. Appendix~\ref{Appx1}, which includes our proof of Schelkunoff's theorem, provides more details on this useful theorem.

\begin{figure}
\centering
\includegraphics[width=1\columnwidth]{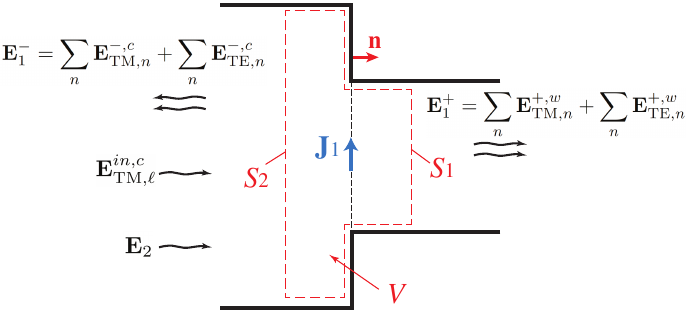}
\caption{An example of step-in discontinuity with a TM incidence. For clarity, only the $E$ fields are explicitly shown in this figure. The integration surfaces on the right and left hand sides of the volume $V$ are denoted by $S_1$ and $S_2$, and are immediately adjacent to each side of the plane $z=0$ (note that the figure is showing exaggerated distances between the surfaces $S_1,S_2$ and $z=0$ for clarity). Here, a single mode is incident ($\mathbf{E}^{in,c}_{\text{TM}\ell}$), a field is reflected back into the waveguide region ($\mathbf{E}^-_1$), and a field is transmitted onto the cavity region ($\mathbf{E}^+_1$), as given by (\ref{E1-}) and (\ref{E1+}). The reflected and transmitted fields can be in both TE and TM in general. Analogous geometries, fields, and surfaces can be constructed for the step-out discontinuity and TE/TM incidences.}
\label{fig_StepIn}
\end{figure}

The general form of {Lorentz's reciprocity theorem \cite{jackson,Collin2} can be formulated by assuming two sets of electric and magnetic current density sources, $(\mathbf{J}_1,\mathbf{M}_1)$ and $(\mathbf{J}_2,\mathbf{M}_2)$, that generate two sets of corresponding fields $(\mathbf{E}_1,\mathbf{H}_1)$ and $(\mathbf{E}_2,\mathbf{H}_2)$, in a volume $V$ enclosed by a surface $S$. Divergence theorem and Maxwell's curl equations, $\nabla\times\mathbf{E}= i\omega\mu\mathbf{B}-\mathbf{M}$ and $\nabla\times\mathbf{H}= -i\omega\epsilon\mathbf{E}+\mathbf{J}$, are then used to yield Lorentz's reciprocity theorem in integral form as
\begin{align}\label{recip}
  &\oint_{S} (\mathbf{E}_2 \times \mathbf{H}_1 -\mathbf{E}_1\times \mathbf{H}_2)\cdot\mathbf{ds}  \nonumber\\
  &=\int_{V}(\mathbf{E}_1\cdot \mathbf{J}_2-\mathbf{H}_1\cdot\mathbf{M}_2-\mathbf{E}_2\cdot \mathbf{J}_1+\mathbf{H}_2\cdot\mathbf{M}_1) dV
  \end{align}
The usefulness of this tool stems from our ability to arbitrarily choose the field sets $(\mathbf{E}_1,\mathbf{H}_1)$ or $(\mathbf{E}_2,\mathbf{H}_2)$ freely, as long as they satisfy Maxwell's equations. Most traditional treatments (e.g.~see \cite{jackson,Collin2,pozar}) tend to analyze simpler scenarios with the first field set $(\mathbf{E}_1,\mathbf{H}_1)$ assumed to be the field expansion in a desired region, while the second field $(\mathbf{E}_2,\mathbf{H}_2)$ is only a single normalized mode from the modes in the full expansion, chosen to be traveling in the backward or forward direction, to obtain the forward ($A^{+}_n, B^{+}_n$) or backward expansion coefficients ($A^{-}_n, B^{-}_n$), respectively \cite{jackson}. This traditional choice, however, will not be valid in the full-scatter analysis in a non-uniform structure, and we utilize a different choice for $(\mathbf{E}_2,\mathbf{H}_2)$, as shown below in Subsections~\ref{Sec: TM on StepIn}--\ref{Sec: TE on StepIn}, where we find the analytical solutions for the four coefficients ($A^{-}_n, B^{-}_n, A^{+}_n, B^{+}_n$) for each of the step-in and step-out discontinuities, under TE or TM incidence. In total, we obtain a set of 16 equations.

Note that the fields written in the following subsections for the full-scatter derivations are the transverse fields (e.g.~$\mathbf{E}\equiv \mathbf{E}_\perp$), unless otherwise stated, and we can always choose $z=0$ locally to be at the plane of the discontinuity under considerations (longitudinal phase evolution will be taken care of later through propagator matrices).

\subsection{Analysis of TM incidence on a step-in discontinuity} \label{Sec: TM on StepIn}

Consider the full-scatter fields resulting from a single TM mode (index $\ell$) that is incident upon a step-in discontinuity, as shown in Figure~\ref{fig_StepIn}. Applying reciprocity theorem and Schelkunoff's theorem to this discontinuity, as demonstrated in Figure~\ref{fig_Scattering}, we can write the transverse fields ($\mathbf{E}_1,\mathbf{H}_1$) at the discontinuity plane ($z=0$) in terms of the combined incident, reflected and transmitted fields.  Specifically, the incident field $(\mathbf{E}_{\text{TM},\ell}^{in,c}, \mathbf{H}_{\text{TM},\ell}^{in,c})$ gives rise to the equivalent current source 
\begin{align}
\mathbf{J}_1 &=+\hat{n}\times\mathbf{H}_0=\hat{n}\times2\mathbf{H}_{\text{TM},\ell}^{in,c}=2B^{in}_{\ell}\hat{n}\times\hat{\mathbf{H}}_{\text{TM},\ell}^{in,c}, \label{J1}
\end{align}
where we have followed the same notation of reciprocity theorem, taking $\mathbf{J}_1$ as the current source equivalent for the discontinuity (aperture) and the first set of fields ($\mathbf{E}_1,\mathbf{H}_1$) as the fields reflected or transmitted by it.

\begin{figure}
\includegraphics[width=0.9\columnwidth]{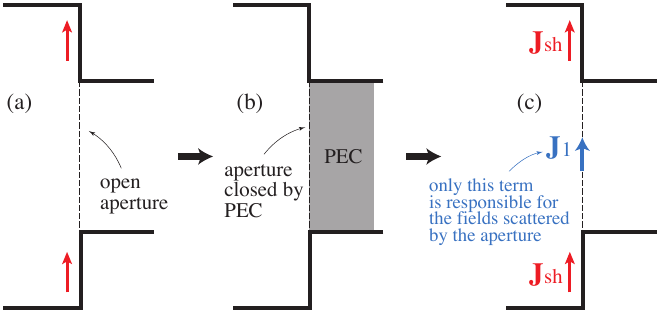}
\caption{When constructing the equivalent problem, by starting from the original problem (a), filling the aperture with a perfect electric conductor (b), and then replacing the aperture by an equivalent current $\mathbf{J}_1$ (c), it is important to distinguish between the surface current $\mathbf{J}_\text{sh.}$ due to the ``shoulders" of the step-in structure, which will always be there, and the current $J_1$ that models only the aperture. Only the latter is the source concerned for generating the scattered fields, $(\mathbf{E}^\pm_1,\mathbf{H}^\pm_1)$, as in Schelkunoff's theorem (see Appendix~\ref{Appx1}).}
\label{fig_Scattering}
\end{figure}

The reflected fields ($\mathbf{E}_1=\mathbf{E}^-_1$, $\mathbf{H}_1=\mathbf{H}^-_1$ in $z<0$) and transmitted fields ($\mathbf{E}_1=\mathbf{E}^+_1$, $\mathbf{H}_1=\mathbf{H}^+_1$ in $z>0$), on the other hand, are given by
\begin{align}
   & \mathbf{E}_1^-=\sum_{n}\mathbf{E}_{\text{TM},n}^{-,c}+\sum_{n}\mathbf{E}_{\text{TE},n}^{-,c}\nonumber \\
    &=\sum_{n}B_n^- \hat{\mathbf{E}}_{\text{TM},n}^{+,c}+\sum_{n}A_n^- \hat{\mathbf{E}}_{\text{TE},n}^{+,c}=\sum_{n}D_n^- \hat{\mathbf{E}}_{n}^{+,c} \label{E1-}\\
    &\mathbf{H}_1^-=\sum_{n}\mathbf{H}_{\text{TM},n}^{-,c}+\sum_{n}\mathbf{H}_{\text{TE},n}^{-,c}\nonumber\\
    & =-\sum_{n}B_n^- \hat{\mathbf{H}}_{\text{TM},n}^{+,c}-\sum_{n}A_n^- \hat{\mathbf{H}}_{\text{TE},n}^{+,c}=-\sum_{n}D_n^- \hat{\mathbf{H}}_{n}^{+,c}\label{H1-}\\
&\mathbf{E}_1^+=\sum_{n}\mathbf{E}_{\text{TM},n}^{+,w}+\sum_{n}\mathbf{E}_{\text{TE},n}^{+,w}\nonumber\\
    &=\sum_{n}B_n^+ \hat{\mathbf{E}}_{\text{TM},n}^{+,w}+\sum_{n}A_n^+ \hat{\mathbf{E}}_{\text{TE},n}^{+,w}=\sum_{n}D_n^+ \hat{\mathbf{E}}_{n}^{+,w}\label{E1+}\\
&\mathbf{H}_1^+=\sum_{n}\mathbf{H}_{\text{TM},n}^{+,w}+\sum_{n}\mathbf{H}_{\text{TE},n}^{+,w}\nonumber\\
    &=-\sum_{n}B_n^+ \hat{\mathbf{H}}_{\text{TM},n}^{+,w}-\sum_{n}A_n^+ \hat{\mathbf{H}}_{\text{TE},n}^{+,w}=-\sum_{n}D_n^+ \hat{\mathbf{H}}_{n}^{+,w}\label{H1+}
\end{align}
where we have made use of (\ref{Er_reduced})--(\ref{Hz_recued}) to reduce all expressions in terms of the forward normalized fields (i.e.~$\hat{\mathbf{E}}_{n}^{+}, \hat{\mathbf{H}}_{n}^{+}$) and, for brevity, used the notation of $\mathbf{E}_n, \mathbf{H}_n$ and $D^\pm_n$ (without explicit TE or TM subscripts) to imply that both $A_n$ and $B_n$ (for TE and TM) are absorbed in the final indexed sum.

Given that the full-scatter fields on each side of the discontinuity are happening in regions of different radii, a careful choice of the field $\mathbf{E}_2$ in reciprocity theorem will allow us to proceed to find all the reflection ($A^-,B^-$) and transmission ($A^+,B^+$) coefficients analytically.  We do so by choosing one of the following \emph{scenarios} for $\mathbf{E}_2$ (it is easy to deduce $\mathbf{H}_2$ from $\mathbf{E}_2$):

(1) To find $B^-_m$, take 
\begin{equation}
\mathbf{E}_2=\begin{cases}
			\hat{\mathbf{E}}_{\text{TM},m}^{+,c}, & z<0\\
            \sum_p \alpha^{+}_{m,p} \hat{\mathbf{E}}_{p}^{+,w}, & z>0
		 \end{cases} \label{Scenario 1}
\end{equation}
         Here, $E_2$ is incident from the cavity region toward the right side, and $\alpha^+_{m,p}$ are some expansion coefficients for the mode in terms of waveguide modes. 
         
(2) To find $A^-_m$, take 
\begin{equation}
\mathbf{E}_2=\begin{cases}
			\hat{\mathbf{E}}_{\text{TE},m}^{+,c}, & z<0\\
            \sum_p \beta^{+}_{m,p} \hat{\mathbf{E}}_{p}^{+,w}, & z>0
		 \end{cases} \label{Scneraio 2}
\end{equation}
         Here, $E_2$ is incident from cavity toward the right side, and $\beta^+_{m,p}$ are some expansion coefficients for the mode in terms of waveguide modes. 

(3) To find $B^+_m$, take 
\begin{equation}
\mathbf{E}_2=\begin{cases}
			\sum_p \gamma^{-}_{m,p} \hat{\mathbf{E}}_{p}^{-,c}, & z<0\\
            \hat{\mathbf{E}}_{\text{TM},m}^{-,w}, & z>0
		 \end{cases} \label{Scenario 3}
\end{equation}
         Here, $E_2$ is incident from the waveguide region toward the left side, and $\gamma^+_{m,p}$ are some expansion coefficients for the mode in terms of cavity modes.

(4) To find $A^+_m$, take 
\begin{equation}
    \mathbf{E}_2=\begin{cases}
			\sum_p \zeta^{-}_{m,p} \hat{\mathbf{E}}_{p}^{-,c}, & z<0\\
            \hat{\mathbf{E}}_{\text{TE},m}^{-,w}, & z>0
		 \end{cases} \label{Scenario 4}
\end{equation}
         Here, $E_2$ is incident from the waveguide region toward the left side, and $\zeta^+_{m,p}$ are some expansion coefficients for the mode in terms of cavity modes.

In the above expressions, $\mathbf{E}_p$ denotes both TE and TM modes, for brevity. The expansion coefficients (such as $\zeta^{-}_{m,p}$, for example) will thus also include those for TE and TM modes. For instance, one can write  $\sum_p \zeta^{-}_{m,p} \hat{\mathbf{E}}_{p}^{-,c}$ more explicitly as $\sum_p B^{-}_{m,p} \hat{\mathbf{E}}_{\text{TM},p}^{-,c}+ \sum_p A^{-}_{m,p} \hat{\mathbf{E}}_{\text{TE},p}^{-,c}$, and so forth. Note that we can derive the analytical expressions for $\alpha^+_{m,p}$, $\beta^+_{m,p}$, $\gamma^+_{m,p}$ and $\zeta^+_{m,p}$ explicitly in the above scenarios, but we do not actually need to, as we will see below. Bearing in mind that $\mathbf{J}_2=0$ and $\mathbf{M}_1=\mathbf{M}_2=0$, reciprocity theorem~(\ref{recip}) now states
\begin{equation}\label{recip_reduced1}
  \oint_{S} (\mathbf{E}_2\times \mathbf{H}_1-\mathbf{E}_1\times \mathbf{H}_2)\cdot\mathbf{ds} =-
  \int_{V}\mathbf{E}_2\cdot \mathbf{J}_1 dV
  \end{equation}

For each of the four scenarios above, we now use (\ref{E1-})--(\ref{H1+}) and (\ref{recip_reduced1}),  to find all $B^-_m, A^-_m, B^+_m, $ and $A^+_m$. 

Consider \emph{Scenario}~(1) to find $B^-_m$, with (\ref{Scenario 1}), we find the LHS (surface integral) of (\ref{recip_reduced1}), bearing in mind that $\oint_S \equiv \int_{S_1} +\int_{S_2}$ (see Figure~\ref{fig_StepIn} for the definition of surfaces $S_1$ and $S_2$), as 
\begin{widetext}
\begin{align}
   \int_{S_2}(\mathbf{E}_2\times \mathbf{H}_1-\mathbf{E}_1\times \mathbf{H}_2)\cdot\mathbf{ds} &= -\int^{2\pi}_0 \int_0^{r_0} r dr d\phi \ \hat{z}\cdot \left[ \hat{\mathbf{E}}^{+,c}_{\text{TM}, m}\times \sum_n D^{-}_{n} \hat{\mathbf{H}}_n^{-,c}-\sum_n D^{-}_{n}\hat{\mathbf{H}}_n^{-,c}\times \hat{\mathbf{E}}^{+,c}_{\text{TM}, m} \right]\nonumber \\
    & =-\sum_n D^{-}_{n}\left[ \iint^{r_0} r dr d\phi \ \hat{z}\cdot \left( \hat{\mathbf{E}}^{+,c}_{\text{TM}, m}\times  \hat{\mathbf{H}}_n^{-,c}\delta_{mn} -\hat{\mathbf{E}}_n^{-,c}\times \hat{\mathbf{H}}^{+,c}_{\text{TM}, m} \delta_{mn}\right) \right]\nonumber\\
    & = -\tilde{B}^{-}_{m}\left[ \iint^{r_0} r dr d\phi \ \hat{z}\cdot \left( -\hat{\mathbf{E}}^{+,c}_{\text{TM}, m}\times  \hat{\mathbf{H}}_{\text{TM},m}^{+,c} -\hat{\mathbf{E}}_{\text{TM},m}^{+,c}\times \hat{\mathbf{H}}^{+,c}_{\text{TM}, m} \right) \right]\nonumber \\
    &= + 2 \tilde{B}^{-}_{m} \underbrace{\iint^{r_0} r dr d\phi \ \hat{z}\cdot (\hat{\mathbf{E}}^{+,c}_{\text{TM}, m}\times  \hat{\mathbf{H}}_{\text{TM},m}^{+,c})}_{=2 \hat{P}^{c}_{\text{TM},m}}= + 4 \tilde{B}^{-}_{m} \hat{P}^{c}_{\text{TM},m} 
\end{align}
\begin{align}
   \int_{S_1}(\mathbf{E}_2\times \mathbf{H}_1-\mathbf{E}_1\times \mathbf{H}_2)\cdot\mathbf{ds} &= +\iint^{r_0} r dr d\phi \ \hat{z}\cdot \left[ \sum_p\alpha^+_{m,p}\hat{\mathbf{E}}^{+,w}_{p}\times \sum_n D^{+}_{n} \hat{\mathbf{H}}_n^{+,w}-\sum D^{+}_{n}\hat{\mathbf{E}}_n^{+,w}\times \sum_p\alpha^+_{m,p}\hat{\mathbf{H}}^{+,w}_{p} \right] \nonumber \\
   &= \sum_{n,p}\alpha^+_{m,p} D^{+}_{n} \iint^{r_0} r dr d\phi \ \hat{z}\cdot \left( \hat{\mathbf{E}}^{+,w}_{p}\times \hat{\mathbf{H}}_n^{+,w}-\hat{\mathbf{E}}_n^{+,w}\times \hat{\mathbf{H}}^{+,w}_{p} \right)=0 \nonumber \\
   \Rightarrow \int_{S_1} +\int_{S_2}&\equiv\oint_{S}(\mathbf{E}_2\times \mathbf{H}_1-\mathbf{E}_1\times \mathbf{H}_2)\cdot\mathbf{ds} = 0+4 \tilde{B}^{-}_{m} \hat{P}^{c}_{\text{TM},m}=4 \tilde{B}^{-}_{m} \hat{P}^{c}_{\text{TM},m}  \label{LHS_StepIn_TM1}
\end{align}
\end{widetext}
where $\tilde{B}^{-}_{m}$ denotes the coefficients of the TM mode scattered by the equivalent source $\mathbf{J}_1$, and $\hat{P}^{c}_{\text{TM},m}$ denotes the complex power carried by the $m$th normalized TM mode from the cavity section side. Namely,
\begin{align}
    \hat{P}^{c}_{\text{TM},\ell} &= \frac{1}{2}\iint^{r_0} rdrd\phi\ \hat{z}\cdot(\hat{\mathbf{E}}^{+,c}_{\text{TM},\ell} \times \hat{\mathbf{H}}^{+,c}_{\text{TM},\ell})\nonumber\\
    &=\pi r_0^2 \left( 1+\frac{\nu^2_{1\ell}}{2k^2 r^2_0}  \right) J^2_0\left(\nu_{1\ell}\right)/(4 Z_0) \label{P_TM_c_l}
\end{align} 

The integrals on surfaces $S_1$ and $S_2$ will carry an opposite sign relative to each other, due to their surface normal vectors being along $\hat{n}$ (shown in Figure~\ref{fig_StepIn}) or opposite to it, respectively.  Note that the appearance of the Kronecher delta function, $\delta_{mn}$, is a direct result of the orthogonality of cavity modes of different type (TE/TM) or same type but different index. The integration domain over the azimuthal angle, $\phi$, is $0$ to $2\pi$ radians, and over the radial direction, $r$, are $0$ to $r_0$. Throughout most of the integrals in this paper, the only limit that will change is the upper limit for the integral in $r$, depending on the region. It will therefore be the only limit explicitly displayed. 

The volume integral in the RHS of (\ref{recip_reduced1}) gives
\begin{widetext}
\begin{align}
    &-\int_V  dV \ \mathbf{E}_2 \cdot \mathbf{J}_1 = -\iiint^a  rdrd\phi dz \left[ \hat{\mathbf{E}}^{+,c}_{\text{TM},m}\cdot \left( 2\hat{z}\times\hat{\mathbf{H}}^{+,c}_{\text{TM},\ell} \right) B^{in}_{\ell} \delta(z)\right]\nonumber \\
    &=+2B^{in}_{\ell}\iint^a r dr d\phi \left( \hat{E_r}^{+,c}_{\text{TM},m}\hat{H_\phi}^{+,c}_{\text{TM},\ell}- \hat{E_\phi}^{+,c}_{\text{TM},m}\hat{H_r}^{+,c}_{\text{TM},\ell}\right)\nonumber\\
    &=+B^{in}_{\ell} \frac{2\pi}{Z_0}\left( 1+\frac{\nu^2_{1\ell}}{2k^2 r^2_0} \right)\int^a dr\left[ r J'_1\left(\frac{\nu_{1m}r}{r_0}\right)J'_1\left(\frac{\nu_{1\ell}r}{r_0}\right) + \frac{r^2_0}{\nu_{1m}\nu_{1\ell} r} J_1\left(\frac{\nu_{1m}r}{r_0}\right)J_1\left(\frac{\nu_{1\ell}r}{r_0}\right)
 \right]\nonumber\\
    &=\frac{\pi B^{in}_{\ell}}{Z_0}\left( 1{+}\frac{\nu^2_{1\ell}}{2k^2 r^2_0} \right) \int^a dr\ r \left[ J_0\left(\frac{\nu_{1m}r}{r_0}\right)J_0\left(\frac{\nu_{1\ell}r}{r_0}\right) {+} J_2\left(\frac{\nu_{1m}r}{r_0}\right)J_2\left(\frac{\nu_{1\ell}r}{r_0}\right)  \right] = \frac{\pi B^{in}_{\ell}}{Z_0}\left( 1{+}\frac{\nu^2_{1\ell}}{2k^2 r^2_0} \right)\begin{cases}
        \Psi_1, & m=\ell\\
        \Psi_2, & m\neq \ell
    \end{cases}   \label{RHS_StepIn_TM1}\\
    & \text{where}\nonumber\\
    \Psi_1&=
    \frac{a^2}{2}\left[ J^2_0\left(\frac{\nu_{1\ell}a}{r_0} \right)+J^2_1\left(\frac{\nu_{1\ell}a}{r_0} \right)+J^2_2\left(\frac{\nu_{1\ell}a}{r_0} \right)-J_1\left(\frac{\nu_{1\ell}a}{r_0} \right)J_3\left(\frac{\nu_{1\ell}a}{r_0} \right) \right] \label{Psi_1}\\
    \Psi_2&=
    \frac{a r_0}{\nu^2_{1\ell}-\nu^2_{1m}}\left\{ \nu_{1\ell} J_1\left(\frac{\nu_{1\ell} a}{r_0} \right) \left[ J_0\left(\frac{\nu_{1m}a}{r_0} \right) -J_2\left(\frac{\nu_{1m}a}{r_0} \right)\right] + \nu_{1m} J_1\left(\frac{\nu_{1m} a}{r_0} \right) \left[ J_2\left(\frac{\nu_{1\ell}a}{r_0} \right) -J_0\left(\frac{\nu_{1\ell}a}{r_0} \right)\right] \right\} \label{Psi_2}
\end{align}
\end{widetext}
Equating (\ref{LHS_StepIn_TM1}) and (\ref{RHS_StepIn_TM1}) gives
\begin{align}
    \tilde{B}^-_m &= \frac{\pi B^{in}_{\ell}}{4 Z_0 \hat{P}^{c}_{\text{TM},m} }\left( 1+\frac{\nu^2_{1\ell}}{2k^2 r^2_0} \right)\begin{cases}
        \Psi_1, & m=\ell\\
        \Psi_2, & m\neq \ell
       \end{cases}
\end{align}
Recall that this reflected field is only due to the equivalent source $\mathbf{J}_1$ and we still need to add it to the $\mathbf{H}_0$ field (produced by the PEC aperture filling), which has $B^-_0=B^{in}_\ell \delta_{m\ell}$ (same magnitude as the incident wave). Thus, we arrive at the total reflection coefficient 
\begin{widetext}
\begin{align}
    B^-_m &= B^-_0 +\tilde{B}^-_m\nonumber\\
    &=B^{in}_\ell \delta_{m\ell} + \tilde{B}^-_m\\
    &= \begin{cases}
        B^{in}_\ell\left\{-1+\frac{\pi a^2 \left( 1+\frac{\nu^2_{1\ell}}{2k^2 r^2_0}  \right)}{8 Z_0 \hat{P}^{c}_{\text{TM},\ell} }\left[ J^2_0\left(\frac{\nu_{1\ell}a}{r_0} \right)+J^2_1\left(\frac{\nu_{1\ell}a}{r_0} \right)+J^2_2\left(\frac{\nu_{1\ell}a}{r_0} \right)-J_1\left(\frac{\nu_{1\ell}a}{r_0} \right)J_3\left(\frac{\nu_{1\ell}a}{r_0} \right) \right] \right\}, & m=\ell\\
        \frac{\pi  B^{in}_\ell a r_0 \left( 1+\frac{\nu^2_{1\ell}}{2k^2 r^2_0}  \right)}{4 Z_0 \hat{P}^{c}_{\text{TM},m} (\nu^2_{1\ell}-\nu^2_{1m})}\left\{ \nu_{1\ell} J_1\left(\frac{\nu_{1\ell} a}{r_0} \right) \left[ J_0\left(\frac{\nu_{1m}a}{r_0} \right) -J_2\left(\frac{\nu_{1m}a}{r_0} \right)\right] + \nu_{1m} J_1\left(\frac{\nu_{1m} a}{r_0} \right) \left[ J_2\left(\frac{\nu_{1\ell}a}{r_0} \right) -J_0\left(\frac{\nu_{1\ell}a}{r_0} \right)\right] \right\}, & m\neq \ell
       \end{cases} \label{B- stepIn1}
\end{align}
\end{widetext}
    
We observe from (\ref{B- stepIn1}) how the reflected mode whose index ($m$) is the same as the incident mode index ($\ell$) will have a different expression compared to modes of different index ($m\neq \ell$). This is expected, since the coupling between the mode and its own reflected wave (same index) will naturally differ from the coupling with others modes, due to the discontinuity.\\

Consider now \emph{Scenario}~(2), to find $A^-_m$, by taking $\mathbf{E}_2$ in (\ref{Scneraio 2}) and applying (\ref{recip_reduced1}) in a similar fashion as the first scenario. We find for the LHS and RHS of (\ref{recip_reduced1}) as 
\begin{align}
   &\oint_{S}(\mathbf{E}_2\times \mathbf{H}_1 -\mathbf{E}_1\times \mathbf{H}_2)\cdot\mathbf{ds}  \nonumber\\
   & \qquad \qquad  = 0+ 4 \tilde{A}^-_m \hat{P}^{c}_{\text{TE},m} = 4 \tilde{A}^-_m \hat{P}^{c}_{\text{TE},m} \label{LHS2}
   \end{align}
   \begin{align}
   &-\int_V  dV \ \mathbf{E}_2 \cdot \mathbf{J}_1 \nonumber\\
   &= -\iiint^a  rdrd\phi dz \left[ \hat{\mathbf{E}}^{+,c}_{\text{TE},m}\cdot \left( 2\hat{z}\times\hat{\mathbf{H}}^{+,c}_{\text{TM},\ell} \right) B^{in}_{\ell} \delta(z)\right]\nonumber\\
    &=2B^{in}_{\ell}\iint^a r dr d\phi \left( \hat{E_r}^{+,c}_{\text{TE},m}\hat{H_\phi}^{+,c}_{\text{TM},\ell}- \hat{E_\phi}^{+,c}_{\text{TE},m}\hat{H_r}^{+,c}_{\text{TM},\ell}\right)\nonumber\\
    &=-B^{in}_{\ell} \frac{2\pi}{Z_0}\left( 1+\frac{\nu^2_{1\ell}}{2k^2 r^2_0} \right)\int^a dr\left[ \frac{r_0}{\nu'_{1m}} J_1\left(\frac{\nu'_{1m}r}{r_0}\right) \right.\nonumber\\
    & \left. \qquad \qquad \  \cdot J'_1\left(\frac{\nu_{1\ell}r}{r_0}\right) + \frac{r_0}{\nu_{1\ell}} J'_1\left(\frac{\nu'_{1m}r}{r_0}\right)J_1\left(\frac{\nu_{1\ell}r}{r_0}\right)
 \right]\nonumber\\
 &=-B^{in}_{\ell} \frac{2\pi}{Z_0}\left( 1+\frac{\nu^2_{1\ell}}{2k^2 r^2_0} \right) \frac{r^2_0 J_1\left(\frac{\nu'_{1m}a}{r_0} \right)J_1\left(\frac{\nu_{1\ell}a}{r_0} \right)}{\nu_{1\ell}\nu'_{1m}}, \label{RHS2}
\end{align}
where $\hat{P}^{c}_{\text{TE},m}$ denotes the complex power carried by the $m$th normalized TE mode from the cavity section side, and is given by
\begin{align}
    \hat{P}^{c}_{\text{TE},\ell} &= \frac{1}{2}\iint^{r_{0}} rdrd\phi\ \hat{z}\cdot(\hat{\mathbf{E}}^{+,c}_{\text{TE},\ell} \times \hat{\mathbf{H}}^{+,c}_{\text{TE},\ell})\label{Power_TE_c_l}\\
    &=\frac{\pi r_{0}^2}{4 Z_0} \left( \frac{\nu'^2_{1\ell}}{2k^2 r_{0}^2} -1 \right) \left( \frac{1}{\nu'^{2}_{1\ell}}-1 \right) J^2_1\left(\nu'_{1\ell}\right) 
\end{align} 

Upon equating (\ref{LHS2}) and (\ref{RHS2}) we obtain
\begin{align}
    \tilde{A}^-_m &= - \frac{\pi B^{in}_{\ell}}{2 Z_0 \hat{P}^{c}_{\text{TE},m}}\left( 1+\frac{\nu^2_{1\ell}}{2k^2 r^2_0} \right) \frac{r^2_0 J_1\left(\frac{\nu'_{1m}a}{r_0} \right)J_1\left(\frac{\nu_{1\ell}a}{r_0} \right)}{\nu_{1\ell}\nu'_{1m}}
\end{align}
Noting that $A^-_0=0$, since the incidence was purely TM in this case, we arrive at the total $A^-_m$
\begin{align}
    &A^-_m= A^-_0 +\tilde{A}^-_m=0 + \tilde{A}^-_m \nonumber\\
    &=- \frac{\pi B^{in}_{\ell} r^2_0 J_1\left(\frac{\nu'_{1m}a}{r_0} \right)J_1\left(\frac{\nu_{1\ell}a}{r_0} \right) \left( 1+\frac{\nu^2_{1\ell}}{2k^2 r^2_0} \right)}{2 Z_0 \hat{P}^{c}_{\text{TE},m} \nu_{1\ell}\nu'_{1m}}  \label{A- stepIn2}
\end{align}

For \emph{Scenario} (3), we find $B^+_m$ by utilizing $\mathbf{E}_2$ in (\ref{Scenario 3}) and calculating the LHS and RHS of (\ref{recip_reduced1}), to yield
\begin{align}
   &\oint_{S}(\mathbf{E}_2\times \mathbf{H}_1-\mathbf{E}_1\times \mathbf{H}_2)\cdot\mathbf{ds} = 0+ 4 B^+_m \hat{P}^{w}_{\text{TM},m} \label{LHS3}\\
   &-\int_V  dV \ \mathbf{E}_2 \cdot \mathbf{J}_1 \nonumber\\
   & \ \ = -\iiint^a  rdrd\phi dz \left[ \hat{\mathbf{E}}^{-,w}_{\text{TM},m}\cdot \left( 2\hat{z}\times\hat{\mathbf{H}}^{+,c}_{\text{TM},\ell} \right) B^{in}_{\ell} \delta(z)\right]\nonumber\\
    &=2B^{in}_{\ell}\iint^a r dr d\phi \left( \hat{E_r}^{+,w}_{\text{TM},m}\hat{H_\phi}^{+,c}_{\text{TM},\ell}{-} \hat{E_\phi}^{+,w}_{\text{TM},m}\hat{H_r}^{+,c}_{\text{TM},\ell}\right)\nonumber\\
    &=B^{in}_{\ell} \frac{2\pi}{Z_0}\left( 1+\frac{\nu^2_{1\ell}}{2k^2 r^2_0} \right)\int^a dr\left[ r J'_1\left(\frac{\nu_{1m}r}{r_0}\right)J'_1\left(\frac{\nu_{1\ell}r}{r_0}\right) \right. \nonumber\\
    & \ \ \ \ \ \ \ \ \ \ \ \ \ \ \left. + \frac{a r_0}{\nu_{1\ell} r} J_1\left(\frac{\nu_{1m}r}{r_0}\right)J_1\left(\frac{\nu_{1\ell}r}{r_0}\right)
 \right]\nonumber\\
 &=B^{in}_{\ell} \frac{2\pi}{Z_0}\left( 1+\frac{\nu^2_{1\ell}}{2k^2 r^2_0} \right) \frac{a \nu_{1\ell} J_0\left(\nu_{1m} \right)J_1\left(\frac{\nu_{1\ell}a}{r_0} \right)}{r_0 (\nu^2_{1\ell}/r^2_0-\nu^2_{1m}/a^2)}\label{RHS3}
\end{align}

Upon equating (\ref{LHS3}) and (\ref{RHS3}), we obtain
\begin{equation}
    B^+_m= \frac{\pi B^{in}_{\ell} a \nu_{1\ell} J_0\left(\nu_{1m} \right)J_1\left(\frac{\nu_{1\ell}a}{r_0} \right) \left( 1+\frac{\nu^2_{1\ell}}{2k^2 r^2_0} \right)}{2 Z_0 \hat{P}^{w}_{\text{TM},m} r_0 (\nu^2_{1\ell}/r^2_0-\nu^2_{1m}/a^2)}\label{B+ stepIn3}
\end{equation}

Similarly, for \emph{Scenario} (4), we find $A^+_m$ by utilizing $\mathbf{E}_2$ in (\ref{Scenario 4}) and calculating the LHS and RHS of (\ref{recip_reduced1}), to yield
\begin{align}
   &\oint_{S}(\mathbf{E}_2\times \mathbf{H}_1-\mathbf{E}_1\times \mathbf{H}_2)\cdot\mathbf{ds} = 0+ 4 A^+_m \hat{P}^{w}_{\text{TE},m} \label{LHS4}\\
   &-\int_V  dV \ \mathbf{E}_2 \cdot \mathbf{J}_1 \nonumber\\
   &= -\iiint^a  rdrd\phi dz \left[ \hat{\mathbf{E}}^{-,w}_{\text{TE},m}\cdot \left( 2\hat{z}\times\hat{\mathbf{H}}^{+,c}_{\text{TM},\ell} \right) B^{in}_{\ell} \delta(z)\right]\nonumber\\
    &=2B^{in}_{\ell}\iint^a r dr d\phi \left( \hat{E_r}^{+,w}_{\text{TE},m}\hat{H_\phi}^{+,c}_{\text{TM},\ell}- \hat{E_\phi}^{+,w}_{\text{TE},m}\hat{H_r}^{+,c}_{\text{TM},\ell}\right)\nonumber\\
    &=-B^{in}_{\ell} \frac{2\pi}{Z_0}\left( 1+\frac{\nu^2_{1\ell}}{2k^2 r^2_0} \right)\int^a dr\left[ \frac{a}{\nu'_{1m}} J_1\left(\frac{\nu'_{1m}r}{a}\right) \cdot \right. \nonumber\\
    & \ \ \ \ \ \ \ \left. J'_1\left(\frac{\nu_{1\ell}r}{r_0}\right) + \frac{r_0}{\nu_{1\ell}} J'_1\left(\frac{\nu'_{1m}r}{a}\right)J_1\left(\frac{\nu_{1\ell}r}{r_0}\right)
 \right] \nonumber \\
 &=-B^{in}_{\ell} \frac{2\pi}{Z_0}\left( 1+\frac{\nu^2_{1\ell}}{2k^2 r^2_0} \right) \frac{a r_0 J_0\left(\nu'_{1m} \right)J_1\left(\frac{\nu_{1\ell}a}{r_0} \right)}{\nu_{1\ell}} \label{RHS4}
\end{align}

Upon equating (\ref{LHS4}) and (\ref{RHS4}), we obtain
\begin{equation}
    A^+_m= - \frac{}{} \frac{\pi B^{in}_{\ell} a r_0 J_0\left(\nu'_{1m} \right)J_1\left(\frac{\nu_{1\ell}a}{r_0} \right) \left( 1+\frac{\nu^2_{1\ell}}{2k^2 r^2_0} \right)}{2 Z_0 \hat{P}^{w}_{\text{TE},m} \nu_{1\ell}} \label{A+ stepIn4}
\end{equation}
where
\begin{align}
    \hat{P}^{w}_{\text{TE},\ell} &= \frac{1}{2}\iint^{a} rdrd\phi\ \hat{z}\cdot(\hat{\mathbf{E}}^{+,w}_{\text{TE},\ell} \times \hat{\mathbf{H}}^{+,w}_{\text{TE},\ell})\label{Power_TE_w_l}\\
    &=\frac{\pi a^2}{4 Z_0} \left( \frac{\nu'^2_{1\ell}}{2k^2 a^2} -1 \right) \left( \frac{1}{\nu'^{2}_{1\ell}}-1 \right) J^2_1\left(\nu'_{1\ell}\right) 
\end{align} 

\begin{figure}
\centering
\includegraphics[width=\columnwidth]{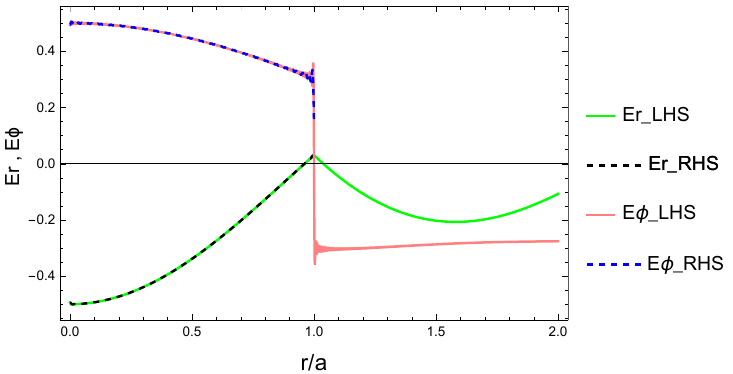}
\caption{Validating the match of each transverse field ($E_r$ and $E_\phi$) on the left-hand side and right-hand side of a step-in discontinuity, for TM incidence. The plot used the results in (\ref{B- stepIn1}), (\ref{A- stepIn2}), (\ref{B+ stepIn3}), and (\ref{A+ stepIn4}) for the field expansions, which were limited to 300 terms and plotted for an example structure that has $r_0/a=2$.}
\label{curves TM_StepIN}
\end{figure}

To verify the correctness of the field expansions on both sides of the discontinuity ($z=0$), we use the coefficients derived in (\ref{B- stepIn1}), (\ref{A- stepIn2}), (\ref{B+ stepIn3}), and (\ref{A+ stepIn4}) in the transverse fields $E_r,E_\phi$, which should match at the $z=0$ boundary. Figure~\ref{curves TM_StepIN} confirms the matching and the correctness of the results for a TM incidence on the step-in discontinuity.

\subsection{Analysis of TM incidence on a step-out discontinuity} \label{Sec: TM on StepOut}

Following similar steps, one can show that, for a TM mode incident on a step-out discontinuity, the reflection coefficients are identically zero. The transmission coefficients for this case are therefore identical for the forward-scatter analysis reported in \cite{NajiTHz2}. To formally prove this, we begin by writing the equivalent current source $\mathbf{J}_1$ for the incident more $(\mathbf{E}_{\text{TM},\ell}^{in,w}, \mathbf{H}_{\text{TM},\ell}^{in,w})$ and the reflected/transmitted ($\mathbf{E}_1,\mathbf{H}_1$) transverse field, to be used in reciprocity theorem. Namely,
\begin{align}    &\mathbf{J}_1=+\hat{n}\times\mathbf{H}_0=\hat{n}\times2\mathbf{H}_{\text{TM},\ell}^{in,w}=2B^{in}_{\ell}\hat{n}\times\hat{\mathbf{H}}_{\text{TM},\ell}^{in,w}\label{J1_TM_stepOut}\\
&\mathbf{E}_1^-=\sum_{n}\mathbf{E}_{\text{TM},n}^{-,w}+\sum_{n}\mathbf{E}_{\text{TE},n}^{-,w}\nonumber\\
    &=\sum_{n}B_n^- \hat{\mathbf{E}}_{\text{TM},n}^{+,w}+\sum_{n}A_n^- \hat{\mathbf{E}}_{\text{TE},n}^{+,w}=\sum_{n}D_n^- \hat{\mathbf{E}}_{n}^{+,w}\\
    &\mathbf{H}_1^-=\sum_{n}\mathbf{H}_{\text{TM},n}^{-,w}+\sum_{n}\mathbf{H}_{\text{TE},n}^{-,w}\nonumber\\
    &=-\sum_{n}B_n^- \hat{\mathbf{H}}_{\text{TM},n}^{+,w}-\sum_{n}A_n^- \hat{\mathbf{H}}_{\text{TE},n}^{+,w}=-\sum_{n}D_n^- \hat{\mathbf{H}}_{n}^{+,w}\label{EH1_TM_stepOut}\\
&\mathbf{E}_1^+=\sum_{n}\mathbf{E}_{\text{TM},n}^{+,c}+\sum_{n}\mathbf{E}_{\text{TE},n}^{+,c}\nonumber\\
 &=\sum_{n}B_n^+ \hat{\mathbf{E}}_{\text{TM},n}^{+,c}+\sum_{n}A_n^+ \hat{\mathbf{E}}_{\text{TE},n}^{+,c}=\sum_{n}D_n^+ \hat{\mathbf{E}}_{n}^{+,c}
\end{align}
\begin{align} 
&\mathbf{H}_1^+=\sum_{n}\mathbf{H}_{\text{TM},n}^{+,c}+\sum_{n}\mathbf{H}_{\text{TE},n}^{+,c}\nonumber\\
    &=-\sum_{n}B_n^+ \hat{\mathbf{H}}_{\text{TM},n}^{+,c}-\sum_{n}A_n^+ \hat{\mathbf{H}}_{\text{TE},n}^{+,c}=-\sum_{n}D_n^+ \hat{\mathbf{H}}_{n}^{+,c}\label{EH2_TM_stepOut}
\end{align}

For the normalized field $\mathbf{E}_2$, we choose from the following four \emph{scenarios}:

(1) to find $B^-_m$, take 
\begin{equation}
    \mathbf{E}_2=\begin{cases}
			\hat{\mathbf{E}}_{\text{TM},m}^{+,w}, & z<0\\
            \sum_p \alpha^{+}_{m,p} \hat{\mathbf{E}}_{p}^{+,c}, & z>0
		 \end{cases} \label{scenario 1_2}
\end{equation}          

(2) To find $A^-_m$, take 
\begin{equation}
    \mathbf{E}_2=\begin{cases}
			\hat{\mathbf{E}}_{\text{TE},m}^{+,w}, & z<0\\
            \sum_p \beta^{+}_{m,p} \hat{\mathbf{E}}_{p}^{+,c}, & z>0
		 \end{cases}\label{scenario 2_2}
\end{equation}

(3) To find $B^+_m$, take 
\begin{equation}
\mathbf{E}_2=\begin{cases}
			\sum_p \gamma^{-}_{m,p} \hat{\mathbf{E}}_{p}^{-,w}, & z<0\\
            \hat{\mathbf{E}}_{\text{TM},m}^{-,c}, & z>0
		 \end{cases}   \label{scenario 3_2}
\end{equation}

(4) To find $A^+_m$, take
\begin{equation}
    \mathbf{E}_2=\begin{cases}
			\sum_p \zeta^{-}_{m,p} \hat{\mathbf{E}}_{p}^{-,w}, & z<0\\
            \hat{\mathbf{E}}_{\text{TE},m}^{-,c}, & z>0
		 \end{cases} \label{scenario4_2}
\end{equation}

Note that, in the above expressions, $\mathbf{E}_p$ denotes both TE and TM modes, for brevity. The expansion coefficients (such as $\zeta^{-}_{m,p}$, for example) will thus include those for TE and TM modes. For example, one can write more explicitly $\sum_p \zeta^{-}_{m,p} \hat{\mathbf{E}}_{p}^{-,c}$ as $\sum_p B^{-}_{m,p} \hat{\mathbf{E}}_{\text{TM},p}^{-,c}+ \sum_p A^{-}_{m,p} \hat{\mathbf{E}}_{\text{TE},p}^{-,c}$, and so forth.

For each of these scenarios, we can now use (\ref{J1_TM_stepOut})--(\ref{EH2_TM_stepOut}) in reciprocity theorem~(\ref{recip_reduced1}) to derive all the field coefficients analytically.

Consider \emph{Scenario} (1) to find $B^-_m$ by taking $\mathbf{E}_2$ from (\ref{scenario 1_2}) in reciprocity theorem (\ref{recip_reduced1}), and bearing in mind that $\oint_S \equiv \int_{S_1} +\int_{S_2}$. We find the LHS as
\begin{align}
   &\int_{S_2}(\mathbf{E}_2\times \mathbf{H}_1-\mathbf{E}_1\times \mathbf{H}_2)\cdot\mathbf{ds} = -\iint^{a} r dr d\phi \nonumber\\
   & \quad \hat{z}\cdot \left[ \hat{\mathbf{E}}^{+,w}_{\text{TM}, m}\times \sum_n D^{-}_{n} \hat{\mathbf{H}}_n^{-,w} -\sum D^{-}_{n}\hat{\mathbf{E}}_n^{-,w}\times \hat{\mathbf{H}}^{+,w}_{\text{TM}, m} \right]\nonumber \\
    & =-\sum_n D^{-}_{n}\left[ \iint^{a} r dr d\phi \ \hat{z}\cdot \left( \hat{\mathbf{E}}^{+,w}_{\text{TM}, m}\times  \hat{\mathbf{H}}_n^{-,w}\delta_{mn} \right. \right. \nonumber\\ 
    & \qquad \qquad \qquad \qquad \left. \left. - \hat{\mathbf{E}}_n^{-,w}\times \hat{\mathbf{H}}^{+,w}_{\text{TM}, m} \delta_{mn}\right) \right]\nonumber\\
    & = -\tilde{B}^{-}_{m}\left[ \iint^{a} r dr d\phi \ \hat{z}\cdot \left( -\hat{\mathbf{E}}^{+,w}_{\text{TM}, m}\times  \hat{\mathbf{H}}_{\text{TM},m}^{+,w} \right. \right. \nonumber\\
    & \qquad \qquad \qquad \qquad  \left. \left. -\hat{\mathbf{E}}_{\text{TM},m}^{+,w}\times \hat{\mathbf{H}}^{+,w}_{\text{TM}, m} \right) \right]\nonumber\\
    &= 2 \tilde{B}^{-}_{m} \iint^{a} r dr d\phi \ \hat{z}\cdot (\hat{\mathbf{E}}^{+,w}_{\text{TM}, m}\times  \hat{\mathbf{H}}_{\text{TM},m}^{+,w})=4 \tilde{B}^{-}_{m} \hat{P}^{w}_{\text{TM},m}
    \end{align}
    \begin{align}
  & \int_{S_1}(\mathbf{E}_2\times \mathbf{H}_1-\mathbf{E}_1\times \mathbf{H}_2)\cdot\mathbf{ds}  \nonumber\\
  &=+ \iint^{r_0} r dr d\phi \ \hat{z}\cdot \left[ \sum_p\alpha^+_{m,p}\hat{\mathbf{E}}^{+,c}_{p}\times \sum_n D^{+}_{n} \hat{\mathbf{H}}_n^{+,c} \right. \nonumber\\
  & \qquad \qquad \qquad  \qquad \qquad \left. -\sum D^{+}_{n}\hat{\mathbf{E}}_n^{+,c}\times \sum_p\alpha^+_{m,p}\hat{\mathbf{H}}^{+,c}_{p} \right]\nonumber 
  \end{align}
\begin{align}
   &= \sum_{n,p}\alpha^+_{m,p} D^{+}_{n} \iint^{r_0} r dr d\phi \ \hat{z}\cdot \left( \hat{\mathbf{E}}^{+,c}_{p}\times \hat{\mathbf{H}}_n^{+,c} \right. \nonumber\\
   & \qquad \qquad \qquad  \qquad \qquad \qquad \qquad \left. -\hat{\mathbf{E}}_n^{+,c}\times \hat{\mathbf{H}}^{+,c}_{p} \right)=0 \nonumber
   \end{align}
   \begin{align}
  \Rightarrow \int_{S_1} +\int_{S_2} &\equiv \oint_{S}(\mathbf{E}_2\times \mathbf{H}_1-\mathbf{E}_1\times \mathbf{H}_2)\cdot\mathbf{ds} \nonumber\\
   & = 0+4 \tilde{B}^{-}_{m} \hat{P}^{w}_{\text{TM},m}\nonumber\\
   & =4 \tilde{B}^{-}_{m} \hat{P}^{w}_{\text{TM},m}  \label{LHS_StepIn_TM_rep1}
\end{align}
On the other hand, the volume integral on the RHS of (\ref{recip_reduced1}) gives
\begin{align}
   & -\int_V  dV \ \mathbf{E}_2 \cdot \mathbf{J}_1 \nonumber  \nonumber\\
    & = -\iiint^a  rdrd\phi dz \left[ \hat{\mathbf{E}}^{+,w}_{\text{TM},m}\cdot \left( 2\hat{z}\times\hat{\mathbf{H}}^{+,w}_{\text{TM},\ell} \right) B^{in}_{\ell} \delta(z)\right]\nonumber\\
    &=2B^{in}_{\ell}\iint^a r dr d\phi \left( \hat{E_r}^{+,w}_{\text{TM},m}\hat{H_\phi}^{+,w}_{\text{TM},\ell}- \hat{E_\phi}^{+,w}_{\text{TM},m}\hat{H_r}^{+,w}_{\text{TM},\ell}\right)\nonumber\\
    &=B^{in}_{\ell} \frac{2\pi}{Z_0}\left( 1+\frac{\nu^2_{1\ell}}{2k^2 a^2} \right)\int^a dr\left[ r J'_1\left(\frac{\nu_{1m}r}{a}\right)J'_1\left(\frac{\nu_{1\ell}r}{a}\right) \right. \nonumber\\
    & \qquad \qquad \qquad \qquad \left. + \frac{a^2}{\nu_{1m}\nu_{1\ell} r} J_1\left(\frac{\nu_{1m}r}{a}\right)J_1\left(\frac{\nu_{1\ell}r}{a}\right)
 \right]\nonumber \\
        &= \frac{\pi B^{in}_{\ell}}{Z_0}\left( 1+\frac{\nu^2_{1\ell}}{2k^2 a^2} \right)\begin{cases}
        a^2\left[ J^2_0\left(\nu_{1\ell}\right) \right], & m=\ell\\
        0, & m\neq \ell
    \end{cases} \label{RHS_StepIn_TM_rep1}
\end{align}
where $\tilde{B}^{-}_{m}$ denotes the coefficients of the TM mode scattered by the equivalent source $\mathbf{J}_1$, and $\hat{P}^{w}_{\text{TM},m}$ denotes the complex power carried by the $m$th normalized TM mode from the waveguide section side. Namely,
\begin{align}
    \hat{P}^{w}_{\text{TM},\ell} &= \frac{1}{2}\iint^a rdrd\phi\ \hat{z}\cdot(\hat{\mathbf{E}}^{+,w}_{\text{TM},\ell} \times \hat{\mathbf{H}}^{+,w}_{\text{TM},\ell})\label{Power_TM_w_l}\\
    &=\frac{\pi a^2}{4 Z_0} \left( 1+\frac{\nu^2_{1\ell}}{2k^2 a^2}  \right) J^2_0\left(\nu_{1\ell}\right) 
\end{align} 

Equating (\ref{LHS_StepIn_TM_rep1}) and (\ref{RHS_StepIn_TM_rep1}) now gives
\begin{align}
    \tilde{B}^-_m &= \frac{\pi B^{in}_{\ell}\left( 1+\frac{\nu^2_{1\ell}}{2k^2 a^2} \right)}{4 Z_0 \hat{P}^{w}_{\text{TM},m} }\begin{cases}
        a^2 J^2_0\left(\nu_{1\ell}\right), & m=\ell\\
        0, & m\neq \ell
       \end{cases}
\end{align}

Recall that this reflected field is only due to the equivalent source $\mathbf{J}_1$. It needs to be added to the field due to the PEC filling, $\mathbf{H}_0$, which has $B^-_0=B^{in}_\ell \delta_{m\ell}$, to give the total reflection. Thus, we arrive at the total reflection coefficient 
\begin{align}
    B^-_m &= B^-_0 +\tilde{B}^-_m=B^{in}_\ell \delta_{m\ell} + \tilde{B}^-_m\nonumber \\
    &= B^{in}_\ell\begin{cases}
        -1+\frac{\pi a^2 J^2_0\left(\nu_{1\ell}\right) \left( 1+\frac{\nu^2_{1\ell}}{2k^2 a^2}  \right)}{4 Z_0 \hat{P}^{w}_{\text{TM},\ell} } =0, & m=\ell\\
        0, & m\neq \ell
       \end{cases} \nonumber\\
       &=0,\label{B- TM stepOut}
\end{align}

Applying \emph{Scenario} (2) to find $A^-_m$, we use (\ref{scenario 2_2}) in (\ref{recip_reduced1}) to yield
\begin{align}
   &\oint_{S}(\mathbf{E}_2\times \mathbf{H}_1-\mathbf{E}_1\times \mathbf{H}_2)\cdot\mathbf{ds} = 0+ 4 \tilde{A}^-_m P^{w}_{\text{TE},m} \label{LHS_s1} \\
   &-\int_V  dV \ \mathbf{E}_2 \cdot \mathbf{J}_1 \nonumber\\
   &=-\iiint^a  rdrd\phi dz \left[ \hat{\mathbf{E}}^{+,w}_{\text{TE},m}\cdot \left( 2\hat{z}\times\hat{\mathbf{H}}^{+,w}_{\text{TM},\ell} \right) B^{in}_{\ell} \delta(z)\right]\nonumber
   \end{align}
   \begin{align}
    &=2B^{in}_{\ell}\iint^a r dr d\phi \left( \hat{E_r}^{+,w}_{\text{TE},m}\hat{H_\phi}^{+,w}_{\text{TM},\ell}- \hat{E_\phi}^{+,w}_{\text{TE},m}\hat{H_r}^{+,w}_{\text{TM},\ell}\right)\nonumber\\
    &=-B^{in}_{\ell} \frac{2\pi}{Z_0}\left( 1+\frac{\nu^2_{1\ell}}{2k^2 a^2} \right)\int^a dr\left[ \frac{a}{\nu'_{1m}} J_1\left(\frac{\nu'_{1m}r}{a}\right)\cdot \right. \nonumber\\
    & \ \ \ \ \ \ \left. J'_1\left(\frac{\nu_{1\ell}r}{a}\right) + \frac{a}{\nu_{1\ell}} J'_1\left(\frac{\nu'_{1m}r}{a}\right)J_1\left(\frac{\nu_{1\ell}r}{a}\right)
 \right]=0, \label{RHS_s1}
\end{align}
since the integral in (\ref{RHS_s1}) is a complete differential.  

Upon equating (\ref{LHS_s1}) and (\ref{RHS_s1}) we obtain $\tilde{A}^-_m = 0$.  Noticing that $A^-_0=0$, since the incidence is purely TM in this case, we then arrive at
\begin{align}
    A^-_m&= A^-_0 +\tilde{A}^-_m=0 + \tilde{A}^-_m= 0 \label{A- TM stepOut}
\end{align}

Applying \emph{Scenario} (3) to find $B^+_m$, we substitute from (\ref{scenario 3_2}) into reciprocity theorem (\ref{recip_reduced1}) to yield
\begin{align}
   &\oint_{S}(\mathbf{E}_2\times \mathbf{H}_1-\mathbf{E}_1\times \mathbf{H}_2)\cdot\mathbf{ds} = 0+ 4 B^+_m \hat{P}^{c}_{\text{TM},m}\label{LHS_s2}\\
   &-\int_V  dV \ \mathbf{E}_2 \cdot \mathbf{J}_1 \nonumber\\
    &= -\iiint^a  rdrd\phi dz \left[ \hat{\mathbf{E}}^{-,c}_{\text{TM},m}\cdot \left( 2\hat{z}\times\hat{\mathbf{H}}^{+,w}_{\text{TM},\ell} \right) B^{in}_{\ell} \delta(z)\right]\nonumber\\
    &=2B^{in}_{\ell}\iint^a r dr d\phi \left( \hat{E_r}^{+,c}_{\text{TM},m}\hat{H_\phi}^{+,w}_{\text{TM},\ell}- \hat{E_\phi}^{+,c}_{\text{TM},m}\hat{H_r}^{+,w}_{\text{TM},\ell}\right)\nonumber\\
    &=B^{in}_{\ell} \frac{2\pi}{Z_0}\left( 1+\frac{\nu^2_{1\ell}}{2k^2 a^2} \right)\int^a dr\left[ r J'_1\left(\frac{\nu_{1m}r}{r_0}\right) \cdot \right. \nonumber\\
    & \ \ \ \ \ \left. J'_1\left(\frac{\nu_{1\ell}r}{r_0}\right) + \frac{a r_0}{\nu_{1\ell} \nu_{1m}r} J_1\left(\frac{\nu_{1m}r}{r_0}\right)J_1\left(\frac{\nu_{1\ell}r}{a}\right)
 \right]\label{temp1}\\
 &=B^{in}_{\ell} \frac{}{}\left( 1+\frac{\nu^2_{1\ell}}{2k^2 a^2} \right) \frac{2\pi a \nu_{1m} J_0(\nu_{1\ell}) J_1\left( \nu_{1m} \frac{a}{r_0} \right)}{Z_0 r_0 \left( \nu_{1m}^2/r_0^2 - \nu_{1\ell}^2/a^2  \right) } \label{RHS_s2}
\end{align}

Upon equating (\ref{LHS_s2}) and (\ref{RHS_s2}) we obtain
\begin{equation}
    B^+_m=\frac{\pi B^{in}_{\ell}}{2 Z_0 \hat P^{c}_{\text{TM},m}}\left( 1+\frac{\nu^2_{1\ell}}{2k^2 a^2} \right) \frac{a \nu_{1m} J_0(\nu_{1\ell}) J_1\left( \nu_{1m} \frac{a}{r_0} \right)}{r_0 (\nu^2_{1m}/r^2_0-\nu^2_{1\ell}/a^2)}. \label{B+ TM stepOut}
\end{equation}

We use \emph{Scenario} (4) to find $A^+_m$, by using (\ref{scenario4_2}) in (\ref{recip_reduced1}) and following the same steps. We find 
\begin{align}
   &\oint_{S}(\mathbf{E}_2\times \mathbf{H}_1-\mathbf{E}_1\times \mathbf{H}_2)\cdot\mathbf{ds} = 0+ 4 A^+_m P^{c}_{\text{TE},m}\label{LHS_s3}\\
   &-\int_V  dV \ \mathbf{E}_2 \cdot \mathbf{J}_1 \nonumber \\
    &= -\iiint^a  rdrd\phi dz \left[ \hat{\mathbf{E}}^{-,c}_{\text{TE},m}\cdot \left( 2\hat{z}\times\hat{\mathbf{H}}^{+,w}_{\text{TM},\ell} \right) B^{in}_{\ell} \delta(z)\right] \nonumber\\
    &=2B^{in}_{\ell}\iint^a r dr d\phi \left( \hat{E_r}^{+,c}_{\text{TE},m}\hat{H_\phi}^{+,w}_{\text{TM},\ell}- \hat{E_\phi}^{+,c}_{\text{TE},m}\hat{H_r}^{+,w}_{\text{TM},\ell}\right)\nonumber\\
    &=-B^{in}_{\ell} \frac{2\pi}{Z_0}\left( 1+\frac{\nu^2_{1\ell}}{2k^2 a^2} \right)\int^a dr\left[ \frac{r_0}{\nu'_{1m}} J_1\left(\frac{\nu'_{1m}r}{r_0}\right) \cdot \right. \nonumber\\
    & \ \ \ \ \ \left. J'_1\left(\frac{\nu_{1\ell}r}{a}\right) + \frac{a}{\nu_{1\ell}} J'_1\left(\frac{\nu'_{1m}r}{r_0}\right)J_1\left(\frac{\nu_{1\ell}r}{a}\right)
 \right]=0, \label{RHS_s3}
\end{align}
since the integral in (\ref{RHS_s3}) is a complete differential.

Upon equating (\ref{LHS_s3}) and (\ref{RHS_s3}) we obtain
\begin{equation}
    A^+_m= 0. \label{A+ TM stepOut}
\end{equation}

\begin{figure}
\centering
\includegraphics[width=\columnwidth]{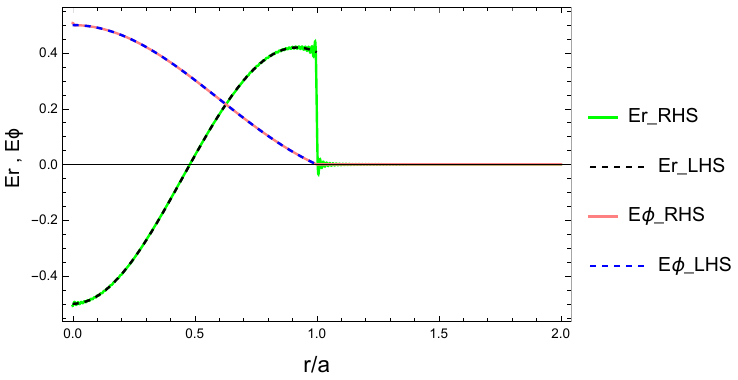}
\caption{Validating the match of each transverse field ($E_r$ and $E_\phi$) on the left-hand side and right-hand side of a step-out discontinuity, for TM incidence. The plot used the results in (\ref{B- TM stepOut}), (\ref{A- TM stepOut}), (\ref{B+ TM stepOut}), and (\ref{A+ TM stepOut}) for the field expansions, which were limited to 300 terms and plotted for an example structure that has $r_0/a=2$.}
\label{curves TM_StepOUT}
\end{figure}
To verify the correctness of the field expansions on both sides of the discontinuity ($z=0$), we use the coefficients derived in (\ref{B- TM stepOut}), (\ref{A- TM stepOut}), (\ref{B+ TM stepOut}), and (\ref{A+ TM stepOut}) in the transverse fields $E_r,E_\phi$, which should match at the $z=0$ boundary. Figure~\ref{curves TM_StepOUT} confirms the matching and the correctness of the results for a TM incidence on the step-in discontinuity.

For brevity, and since the same approach is reused to find the analysis for the TE incidence on step-in and step-out discontuinity, Subsections~\ref{Sec: TE on StepIn} and \ref{Sec: TE on StepOut} will mainly summarize the key field equations and the resulting coefficients. 

\subsection{Analysis of TE incidence on a step-in discontinuity} \label{Sec: TE on StepIn}
Following similar steps for the case with TE incidence, $(\mathbf{E}_{\text{TE},\ell}^{in,c}, \mathbf{H}_{\text{TE},\ell}^{in,c})$, on a step-in discontinuity, we can write the equivalent current source $\mathbf{J}_1$ and ($\mathbf{E}_1, \mathbf{H}_1$) fields as
\begin{align}
\mathbf{J}_1&=+\hat{n}\times\mathbf{H}_0=+\hat{n}\times2\mathbf{H}_{\text{TE},\ell}^{in,c}=2A^{in}_{\ell}\hat{n}\times\hat{\mathbf{H}}_{\text{TE},\ell}^{in,c}\label{J1 TE stepIn}\\
\mathbf{E}_1^-&=\sum_{n}\mathbf{E}_{\text{TM},n}^{-,c}+\sum_{n}\mathbf{E}_{\text{TE},n}^{-,c}\nonumber\\
&=\sum_{n}B_n^- \hat{\mathbf{E}}_{\text{TM},n}^{+,c}+\sum_{n}A_n^- \hat{\mathbf{E}}_{\text{TE},n}^{+,c}=\sum_{n}D_n^- \hat{\mathbf{E}}_{n}^{+,c}\nonumber\\
\mathbf{H}_1^-&=\sum_{n}\mathbf{H}_{\text{TM},n}^{-,c}+\sum_{n}\mathbf{H}_{\text{TE},n}^{-,c}\nonumber\\
&=-\sum_{n}B_n^- \hat{\mathbf{H}}_{\text{TM},n}^{+,c}-\sum_{n}A_n^- \hat{\mathbf{H}}_{\text{TE},n}^{+,c}=-\sum_{n}D_n^- \hat{\mathbf{H}}_{n}^{+,c}\label{EH1- TE stepIn}\\
\mathbf{E}_1^+&=\sum_{n}\mathbf{E}_{\text{TM},n}^{+,w}+\sum_{n}\mathbf{E}_{\text{TE},n}^{+,w}\nonumber\\
&=\sum_{n}B_n^+ \hat{\mathbf{E}}_{\text{TM},n}^{+,w}+\sum_{n}A_n^+ \hat{\mathbf{E}}_{\text{TE},n}^{+,w}=\sum_{n}D_n^+ \hat{\mathbf{E}}_{n}^{+,w}\nonumber\\
\mathbf{H}_1^+&=\sum_{n}\mathbf{H}_{\text{TM},n}^{+,w}+\sum_{n}\mathbf{H}_{\text{TE},n}^{+,w}\nonumber\\
&=-\sum_{n}B_n^+ \hat{\mathbf{H}}_{\text{TM},n}^{+,w}-\sum_{n}A_n^+ \hat{\mathbf{H}}_{\text{TE},n}^{+,w}=-\sum_{n}D_n^+ \hat{\mathbf{H}}_{n}^{+,w}\label{EH1+ TE stepIn}
\end{align}

The normalized field $\mathbf{E}_2$ is chosen out of the following four scenarios, to find the four coefficient sets:

(1) To find $B^-_m$, take 
\begin{equation}
    \mathbf{E}_2=\begin{cases}
			\hat{\mathbf{E}}_{\text{TM},m}^{+,c}, & z<0\\
            \sum_p \alpha^{+}_{m,p} \hat{\mathbf{E}}_{p}^{+,w}, & z>0
		 \end{cases} \label{scenario 1_3}
\end{equation}
         
(2) To find $A^-_m$, take 
\begin{equation}
    \mathbf{E}_2=\begin{cases}
			\hat{\mathbf{E}}_{\text{TE},m}^{+,c}, & z<0\\
            \sum_p \beta^{+}_{m,p} \hat{\mathbf{E}}_{p}^{+,w}, & z>0
		 \end{cases}\label{scenario 2_3}
\end{equation}

(3) To find $B^+_m$, take
\begin{equation}
\mathbf{E}_2=\begin{cases}
			\sum_p \gamma^{-}_{m,p} \hat{\mathbf{E}}_{p}^{-,c}, & z<0\\
            \hat{\mathbf{E}}_{\text{TM},m}^{-,w}, & z>0
		 \end{cases}    \label{scenario 3_3}
\end{equation}
         
(4) To find $A^+_m$, take 
\begin{equation}
    \mathbf{E}_2=\begin{cases}
			\sum_p \zeta^{-}_{m,p} \hat{\mathbf{E}}_{p}^{-,c}, & z<0\\
            \hat{\mathbf{E}}_{\text{TE},m}^{-,w}, & z>0
		 \end{cases}\label{scenario 4_3}
\end{equation}

Using~(\ref{J1 TE stepIn})--(\ref{EH1+ TE stepIn}) and scenarios (\ref{scenario 1_3})--(\ref{scenario 4_3}) in reciprocity theorem~(\ref{recip_reduced1}), we arrive, after mathematical reduction,  at the following coefficients
\begin{align}
    B^-_m &= \frac{\pi A^{in}_{\ell} r_0^2 J_1(\nu_{1m} \frac{a}{r_0}) J_1\left( \nu'_{1\ell} \frac{a}{r_0} \right) \left(\frac{\nu'^{2}_{1l}}{2k^{2}r_0^{2}}-1 \right)}{2 Z_0 \hat{P}^{c}_{\text{TM},m} \nu'_{1\ell}\nu_{1m}} \label{B- TE stepIn}
\end{align}
\begin{widetext}
\begin{align}
    A^-_m &= A^-_0 +\tilde{A}^-_m=A^{in}_\ell \delta_{m\ell} + \tilde{A}^-_m=A^{in}_{\ell}\begin{cases}
        -1 + \frac{-\pi}{4 Z_0 \hat{P}^{c}_{\text{TE},m} }\left(\frac{\nu'^{2}_{1l}}{2k^{2}r_0^{2}}-1 \right)\hat{\Psi}_1, & m=\ell\\
        \frac{-\pi}{4 Z_0 \hat{P}^{c}_{\text{TE},m} }\left(\frac{\nu'^{2}_{1l}}{2k^{2}r_0^{2}}-1 \right)\hat{\Psi}_2, & m\neq \ell
       \end{cases}, \ \text{where } \label{A- TE stepIn}\\
    \hat{\Psi}_1&=
    \frac{a^2}{2}\left[ J^2_0\left(\frac{\nu'_{1\ell}a}{r_0} \right)+J^2_1\left(\frac{\nu'_{1\ell}a}{r_0} \right)+J^2_2\left(\frac{\nu'_{1\ell}a}{r_0} \right)-J_1\left(\frac{\nu'_{1\ell}a}{r_0} \right)J_3\left(\frac{\nu'_{1\ell}a}{r_0} \right) \right]\label{hat Psi 1}\\
    \hat{\Psi}_2&=
    \frac{a r_0}{\nu'^2_{1\ell}-\nu'^2_{1m}}\left\{ \nu'_{1\ell} J_1\left(\frac{\nu'_{1\ell} a}{r_0} \right) \left[ J_0\left(\frac{\nu'_{1m}a}{r_0} \right) -J_2\left(\frac{\nu'_{1m}a}{r_0} \right)\right] + \nu'_{1m} J_1\left(\frac{\nu'_{1m} a}{r_0} \right) \left[ J_2\left(\frac{\nu'_{1\ell}a}{r_0} \right) -J_0\left(\frac{\nu'_{1\ell}a}{r_0} \right)\right] \right\}\label{hat Psi 2}
\end{align}
\end{widetext}
\begin{align}
    B^+_m&= 0 \label{B+ TE stepIn}\\
    A^+_m&= \frac{-\pi A^{in}_{\ell} r_0^2 \nu'_{1m} J_1(\nu'_{1m}) J'_1\left( \nu'_{1\ell} \frac{a}{r_0} \right) \left(\frac{\nu'^{2}_{1l}}{2k^{2}r_0^{2}}-1 \right)}{2Z_0 P^{w}_{\text{TE},m}(\nu'^2_{1m}r^2_0/a^2-\nu'^2_{1\ell})}  \label{A+ TE stepIn}
\end{align}

\begin{figure}
\centering
\includegraphics[width=\columnwidth]{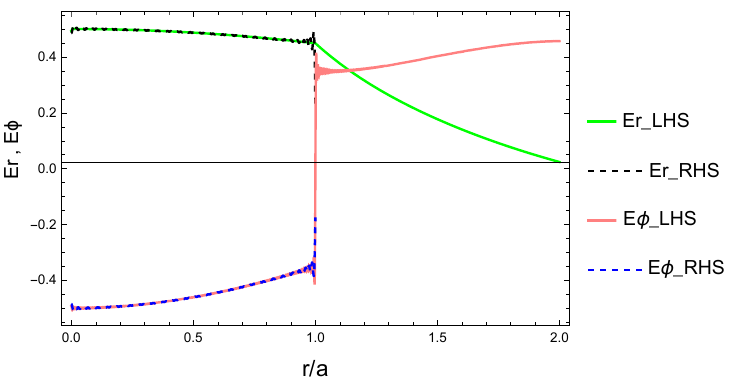}
\caption{Validating the match of each transverse field ($E_r$ and $E_\phi$) on the left-hand side and right-hand side of a step-in discontinuity, for TE incidence. The plot used the results in (\ref{B- TE stepIn}), (\ref{A- TE stepIn}), (\ref{B+ TE stepIn}), and (\ref{A+ TE stepIn}) for the field expansions, which were limited to 300 terms and plotted for an example structure that has $r_0/a=2$.}
\label{curves TE_StepIN}
\end{figure}
To verify the correctness of the field expansions on both sides of the discontinuity ($z=0$), we use the coefficients derived in (\ref{B- TE stepIn}), (\ref{A- TE stepIn}), (\ref{B+ TE stepIn}), and (\ref{A+ TE stepIn}) in the transverse fields $E_r,E_\phi$, which should match at the $z=0$ boundary. Figure~\ref{curves TE_StepIN} confirms the matching and the correctness of the results for a TE incidence on the step-in discontinuity.

\subsection{Analysis of TE incidence on a step-out discontinuity} \label{Sec: TE on StepOut}

Conducting the same analysis steps for a TE incidence, $(\mathbf{E}_{\text{TE},\ell}^{in,c},\mathbf{H}_{\text{TE},\ell}^{in,c})$, on a step-out will lead to reflection coefficients are identically zero. The transmission coefficients for this case are therefore identical for the forward-scatter analysis reported in \cite{NajiTHz2}. We can write the equivalent current source $\mathbf{J}_1$ and ($\mathbf{E}_1, \mathbf{H}_1$) fields as
\begin{align}
\mathbf{J}_1&=+\hat{n}\times\mathbf{H}_0=\hat{n}\times2\mathbf{H}_{\text{TE},\ell}^{in,w}=2A^{in}_{\ell}\hat{n}\times\hat{\mathbf{H}}_{\text{TE},\ell}^{in,w}\label{J1 TE stepOut}
\end{align}
\begin{align}
\mathbf{E}_1^-&=\sum_{n}\mathbf{E}_{\text{TM},n}^{-,w}+\sum_{n}\mathbf{E}_{\text{TE},n}^{-,w}\nonumber\\
&=\sum_{n}B_n^- \hat{\mathbf{E}}_{\text{TM},n}^{+,w}+\sum_{n}A_n^- \hat{\mathbf{E}}_{\text{TE},n}^{+,w}=\sum_{n}D_n^- \hat{\mathbf{E}}_{n}^{+,w}\nonumber\\
\mathbf{H}_1^-&=\sum_{n}\mathbf{H}_{\text{TM},n}^{-,w}+\sum_{n}\mathbf{H}_{\text{TE},n}^{-,w}\nonumber\\
&=-\sum_{n}B_n^- \hat{\mathbf{H}}_{\text{TM},n}^{+,w}-\sum_{n}A_n^- \hat{\mathbf{H}}_{\text{TE},n}^{+,w}=-\sum_{n}D_n^- \hat{\mathbf{H}}_{n}^{+,w}\label{EH1- TE stepOut}\\
\mathbf{E}_1^+&=\sum_{n}\mathbf{E}_{\text{TM},n}^{+,c}+\sum_{n}\mathbf{E}_{\text{TE},n}^{+,c}\nonumber\\
&=\sum_{n}B_n^+ \hat{\mathbf{E}}_{\text{TM},n}^{+,c}+\sum_{n}A_n^+ \hat{\mathbf{E}}_{\text{TE},n}^{+,c}=\sum_{n}D_n^+ \hat{\mathbf{E}}_{n}^{+,c}\nonumber\\
\mathbf{H}_1^+&=\sum_{n}\mathbf{H}_{\text{TM},n}^{+,c}+\sum_{n}\mathbf{H}_{\text{TE},n}^{+,c}\nonumber\\
&=-\sum_{n}B_n^+ \hat{\mathbf{H}}_{\text{TM},n}^{+,c}-\sum_{n}A_n^+ \hat{\mathbf{H}}_{\text{TE},n}^{+,c}=-\sum_{n}D_n^+ \hat{\mathbf{H}}_{n}^{+,c}\label{EH1+ TE stepOut}
\end{align}

The normalized field $\mathbf{E}_2$ is chosen out of the following four scenarios, to find the four coefficient sets:

(1) To find $B^-_m$, take
\begin{equation}
\mathbf{E}_2=\begin{cases}
			\hat{\mathbf{E}}_{\text{TM},m}^{+,w}, & z<0\\
            \sum_p \alpha^{+}_{m,p} \hat{\mathbf{E}}_{p}^{+,c}, & z>0
		 \end{cases}    
\end{equation} \label{scenario 1_4}

(2) To find $A^-_m$, take 
\begin{equation}
    \mathbf{E}_2=\begin{cases}
			\hat{\mathbf{E}}_{\text{TE},m}^{+,w}, & z<0\\
            \sum_p \beta^{+}_{m,p} \hat{\mathbf{E}}_{p}^{+,c}, & z>0
		 \end{cases}\label{scenario 2_4}
\end{equation}
         
(3) To find $B^+_m$, take
\begin{equation}
    \mathbf{E}_2=\begin{cases}
			\sum_p \gamma^{-}_{m,p} \hat{\mathbf{E}}_{p}^{-,w}, & z<0\\
            \hat{\mathbf{E}}_{\text{TM},m}^{-,c}, & z>0
		 \end{cases} \label{scenario 3_4}
\end{equation}
         
(4) To find $A^+_m$, take
\begin{equation}
    \mathbf{E}_2=\begin{cases}
			\sum_p \zeta^{-}_{m,p} \hat{\mathbf{E}}_{p}^{-,w}, & z<0\\
            \hat{\mathbf{E}}_{\text{TE},m}^{-,c}, & z>0
		 \end{cases} \label{scenario 4_4}
\end{equation}

Using~(\ref{J1 TE stepOut})--(\ref{EH1+ TE stepOut}) and scenarios (\ref{scenario 1_4})--(\ref{scenario 4_4}) in reciprocity theorem~(\ref{recip_reduced1}), we arrive, after mathematical reduction,  at the following coefficients
\begin{align}
    B^-_m &= 0 \label{B- TE stepOut} \\
    A^-_m &= 0 \label{A- TE stepOut}\\
    B^+_m &= \frac{\pi A^{in}_{\ell} a r_0 J_{1}\left( \frac{\nu_{1m}a}{r_0}\right)J_{1}(\nu'_{1\ell}) \left(\frac{\nu'^{2}_{1l}}{2k^{2}a^{2}}-1 \right)}{2Z_0 \hat{P}^{c}_{\text{TM},m} \nu'_{1\ell} \nu_{1m}}\label{B+ TE stepOut}\\
    A^+_m&= \frac{\pi A^{in}_{\ell}r_0^2 \nu'_{1\ell} J_1(\nu'_{1\ell}) J'_1\left( \nu'_{1m} \frac{a}{r_0} \right) \left(1-\frac{\nu'^{2}_{1l}}{2k^{2}a^{2}} \right)}{2 Z_0 \hat{P}^{c}_{\text{TE},m} (\nu'^2_{1\ell}r^2_0/a^2-\nu'^2_{1m})}  \label{A+ TE stepOut}
\end{align}

\begin{figure}
\centering
\includegraphics[width=\columnwidth]{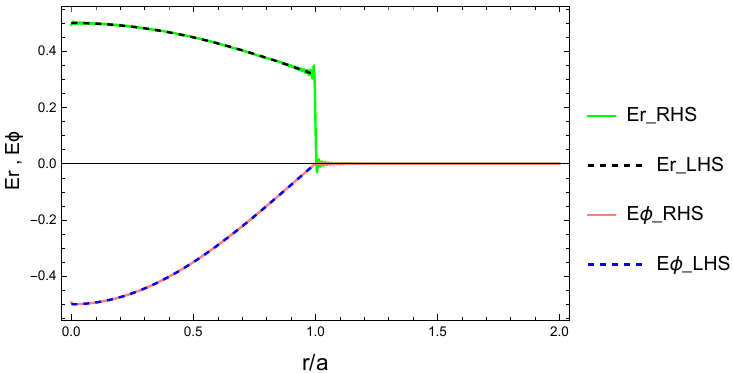}
\caption{Validating the match of each transverse field ($E_r$ and $E_\phi$) on the left-hand side and right-hand side of a step-out discontinuity, for TE incidence. The plot used the results in (\ref{B- TE stepOut}), (\ref{A- TE stepOut}), (\ref{B+ TE stepOut}), and (\ref{A+ TE stepOut}) for the field expansions, which were limited to 300 terms and plotted for an example structure that has $r_0/a=2$.}
\label{curves TE_StepOUT}
\end{figure}
To verify the correctness of the field expansions on both sides of the discontinuity ($z=0$), we use the coefficients derived in (\ref{B- TE stepOut}), (\ref{A- TE stepOut}), (\ref{B+ TE stepOut}), and (\ref{A+ TE stepOut}) in the transverse fields $E_r,E_\phi$, which should match at the $z=0$ boundary. Figure~\ref{curves TE_StepOUT} confirms the matching and the correctness of the results for a TE incidence on the step-out discontinuity.
\begin{figure}
	\includegraphics[width=0.75\columnwidth]{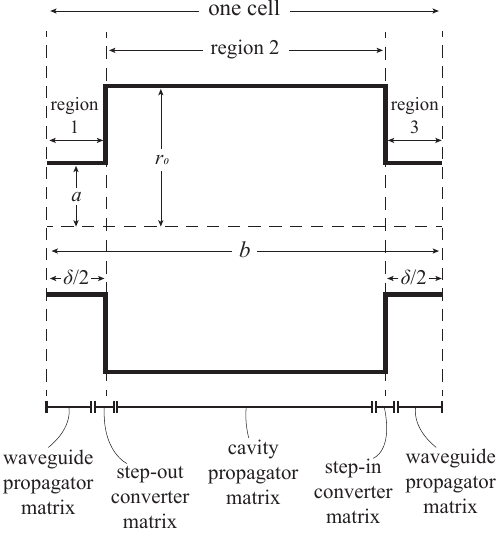}
	\caption{A sketch illustrating a cell in the iris line, and how the different sections of the cell are modelled using different transmission matrices in the numerical analysis.}
	\label{fig: cell}
\end{figure}

\begin{figure*}
	\includegraphics[width=0.9\textwidth]{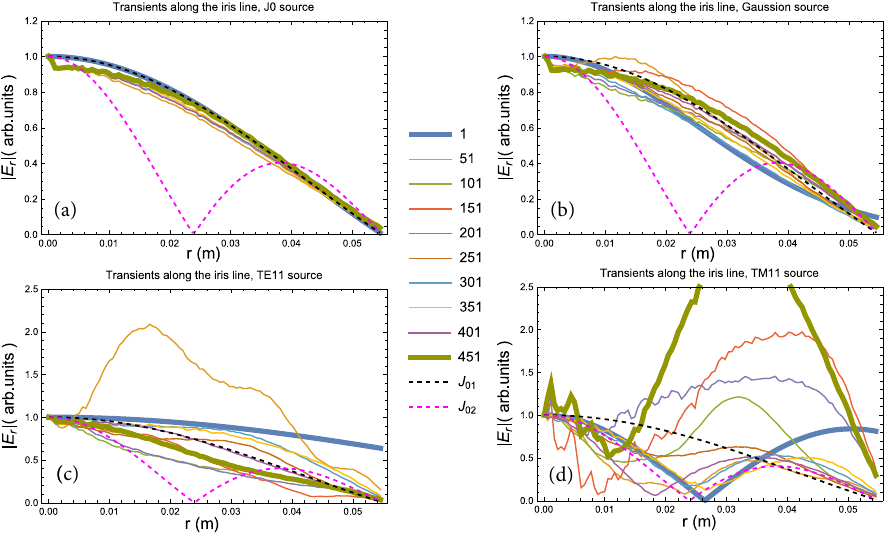}
	\caption{Normalized $E_r$ field transients evolving along the iris line toward steady state. For clarity, the fields are sampled at intervals of 50 irises, on the iris numbers shown in the middle of the figure (the heavy blue trace indicates the first iris, and the heavy green trace represents that last iris). Also shows are the ideal (reference) curves of the first two $J_0$ modes (at the first and second zeros of $J_0$). Different excitation scenarios are shown: (a) the well-matched $J_0(2.4r/a)$ source, (b) the backed-off Gaussian source, (c) a TE$_{11}$ source, and (d) a TM$_{11}$ source. See further discussion in the text.}
	\label{fig: allPlots}
\end{figure*}

\subsection{Summary of full-scatter coefficients}
The results obtained in Sebsections~\ref{Sec: TM on StepIn}--\ref{Sec: TE on StepOut} can now be summarized as follows, where the acronyms ``SI" and ``SO" stand for step-in and step-out, respectively.
\begin{equation}
    B^-_m=\begin{cases} \label{B- summary}
        \frac{\pi B^{in}_{\ell}\left( 1+\frac{\nu^2_{1\ell}}{2k^2 r^2_0} \right)}{4 Z_0 \hat{P}^{c}_{\text{TM},m} }\begin{cases}
        \Psi_1, & m=\ell\\
        \Psi_2, & m\neq \ell
       \end{cases}, & \text{TM$_\ell^\text{SI}$}\\
        0, & \text{TM$_\ell^\text{SO}$}\\
        \frac{\pi A^{in}_{\ell} r_0^2 J_1(\nu_{1m} \frac{a}{r_0}) J_1\left( \nu'_{1\ell} \frac{a}{r_0} \right) \left(\frac{\nu'^{2}_{1l}}{2k^{2}r_0^{2}}-1 \right)}{2 Z_0 \hat{P}^{c}_{\text{TM},m} \nu'_{1\ell}\nu_{1m}}, & \text{TE$_\ell^\text{SI}$}\\
        0, & \text{TE$_\ell^\text{SO}$}\\
    \end{cases}
    \end{equation}    
\begin{equation}
    A^-_m=\begin{cases} \label{A- summary}
        - \frac{\pi B^{in}_{\ell} r^2_0 J_1\left(\frac{\nu'_{1m}a}{r_0} \right)J_1\left(\frac{\nu_{1\ell}a}{r_0} \right) \left( 1+\frac{\nu^2_{1\ell}}{2k^2 r^2_0} \right)}{2 Z_0 \hat{P}^{c}_{\text{TE},m} \nu_{1\ell}\nu'_{1m}}, & \text{TM$_\ell^\text{SI}$}\\
        0, & \text{TM$_\ell^\text{SO}$}\\
        \begin{cases}
        -A^{in}_{\ell} + \frac{-\pi A^{in}_{\ell} \left(\frac{\nu'^{2}_{1l}}{2k^{2}r_0^{2}}-1 \right)}{4 Z_0 \hat{P}^{c}_{\text{TE},m} }\hat{\Psi}_1, & m=\ell\\
        \frac{-\pi A^{in}_{\ell} \left(\frac{\nu'^{2}_{1l}}{2k^{2}r_0^{2}}-1 \right)}{4 Z_0 \hat{P}^{c}_{\text{TE},m} }\hat{\Psi}_2, & m\neq \ell
       \end{cases}, & \text{TE$_\ell^\text{SI}$}\\
        0, & \text{TE$_\ell^\text{SO}$}\\
    \end{cases}
\end{equation}    
\begin{equation}
    B^+_m=\begin{cases} \label{B+summary}
       \frac{\pi B^{in}_{\ell} a \nu_{1\ell} J_0\left(\nu_{1m} \right)J_1\left(\frac{\nu_{1\ell}a}{r_0} \right) \left( 1+\frac{\nu^2_{1\ell}}{2k^2 r^2_0} \right)}{2 Z_0 \hat{P}^{w}_{\text{TM},m} r_0 (\nu^2_{1\ell}/r^2_0-\nu^2_{1m}/a^2)}, & \text{TM$_\ell^\text{SI}$}\\
        \frac{\pi B^{in}_{\ell} a \nu_{1m} J_0(\nu_{1\ell}) J_1\left( \nu_{1m} \frac{a}{r_0} \right) \left( 1+\frac{\nu^2_{1\ell}}{2k^2 a^2} \right)}{2 Z_0 \hat P^{c}_{\text{TM},m} r_0 (\nu^2_{1m}/r^2_0-\nu^2_{1\ell}/a^2)}, & \text{TM$_\ell^\text{SO}$}\\
       0, & \text{TE$_\ell^\text{SI}$}\\
        \frac{\pi A^{in}_{\ell} a r_0 J_{1}\left( \frac{\nu_{1m}a}{r_0}\right)J_{1}(\nu'_{1\ell}) \left(\frac{\nu'^{2}_{1l}}{2k^{2}a^{2}}-1 \right)}{2Z_0 \hat{P}^{c}_{\text{TM},m} \nu'_{1\ell} \nu_{1m}}, & \text{TE$_\ell^\text{SO}$}\end{cases}
\end{equation}    
\begin{equation}
    A^+_m=\begin{cases} \label{A+ summary}
       - \frac{}{} \frac{\pi B^{in}_{\ell} a r_0 J_0\left(\nu'_{1m} \right)J_1\left(\frac{\nu_{1\ell}a}{r_0} \right) \left( 1+\frac{\nu^2_{1\ell}}{2k^2 r^2_0} \right)}{2 Z_0 \hat{P}^{w}_{\text{TE},m} \nu_{1\ell}}, & \text{TM$_\ell^\text{SI}$}\\
        0, & \text{TM$_\ell^\text{SO}$}\\
       \frac{-\pi A^{in}_{\ell} r_0^2 \nu'_{1m} J_1(\nu'_{1m}) J'_1\left( \nu'_{1\ell} \frac{a}{r_0} \right) \left(\frac{\nu'^{2}_{1l}}{2k^{2}r_0^{2}}-1 \right)}{2Z_0 P^{w}_{\text{TE},m}(\nu'^2_{1m}r^2_0/a^2-\nu'^2_{1\ell})}, & \text{TE$_\ell^\text{SI}$}\\
        \frac{\pi A^{in}_{\ell}r_0^2 \nu'_{1\ell} J_1(\nu'_{1\ell}) J'_1\left( \nu'_{1m} \frac{a}{r_0} \right) \left(1-\frac{\nu'^{2}_{1l}}{2k^{2}a^{2}} \right)}{2 Z_0 \hat{P}^{c}_{\text{TE},m} (\nu'^2_{1\ell}r^2_0/a^2-\nu'^2_{1m})}, & \text{TE$_\ell^\text{SO}$}
    \end{cases}
\end{equation}    
where $\Psi_{1,2}$ are given in given by (\ref{Psi_1}) and (\ref{Psi_2}), $\hat{\Psi}_{1,2}$ are given in given by (\ref{hat Psi 1}) and (\ref{hat Psi 2}), and $\hat{P}^{c}_{\text{TM},m}$, $\hat{P}^{c}_{\text{TE},m}$, $\hat{P}^{w}_{\text{TM},m}$, $\hat{P}^{w}_{\text{TE},m}$ are given, respectively, by (\ref{P_TM_c_l}), (\ref{Power_TE_c_l}), (\ref{Power_TM_w_l}) and (\ref{Power_TE_w_l})

\section{Implementation and Numerical Examples} \label{sec: implementations}
The theoretical findings for the full-scatter analysis developed in Section~\ref{Sec: analysis} are now implemented on a computer, to solve for numerical examples and to benchmark the results against other known models. The iris line is divided into cells, as shown in Figure~\ref{fig: cell} for example, where each cell is represented by a cell transmission matrix $\mathsf{C}$ and the entire line is made up by cascading a number of such cells \cite{NajiTHz2}. The total scattering matrix of the line is therefore the multiplication of multiple cell matrices, in their order of occurrence. As described in \cite{NajiTHz2}, each cell matrix is found by modeling the cell by 5 submatrices, representing a propagator matrix ($\mathsf{P}_{1}$) for waveguide region~1, followed by a step-out scattering matrix ($\mathsf{S}_{\text{step-out}}$), a propagator matrix ($\mathsf{P}_{2}$) in cavity region~2, a step-in scattering matrix ($\mathsf{S}_{\text{step-in}}$), and finally a propagator matrix ($\mathsf{P}_{3}$) in  waveguide region~3; see Figure~\ref{fig: cell}. As an example, the output vector matrix $\mathsf{E}^{\mathsf{T}}_{r,\text{out}}$, representing the $E_r$ field TE and TM modes scattered by a cell, for a given input vector $\mathsf{E}^{\mathsf{T}}_{r,\text{out}}$, is given by
\begin{equation}
    \mathsf{E}^{\mathsf{T}}_{r,\text{out}}=\mathsf{E}^{\mathsf{T}}_{r,\text{in}}\cdot \underbrace{\mathsf{P}_{1} \cdot \mathsf{S}_{\text{step-out}} \cdot \mathsf{P}_{2} \cdot \mathsf{S}_{\text{step-in}} \cdot \mathsf{P}_{3}}_{\equiv \ \mathsf{C}},
\end{equation}
where the superscript $\mathsf{T}$ denoted the matrix transpose. The scattering (or ``converter") matrices are populated with the coefficients we derived in Section~\ref{Sec: analysis} and summarized in (\ref{B- summary})--(\ref{A+ summary}). For full details on the construction of these matrices, the reader is referred to reference \cite{NajiTHz2}.

This scattering formulation has a higher computational efficiency compared to standard finite-elements methods (FEM) or the eigensolution method developed in \cite{NajiTHz1}, for the overmoded structures under considerations. In addition to lending itself to structures that are not fully periodic (e.g.~mechanical defects representing slightly different matrices), this method  allows us to observe the extent of the transient regime that happens at the entrance of the iris line, as we excite the structure by a given source. In this Section we benchmark the results of this full-scatter analysis against the Vainstein limit for thin screens ($\delta\rightarrow 0$) \cite{DESY,Vainstein1,Vainstein2,Vainstein3}, the eigensolution method \cite{NajiTHz1}, and the high-frequency forward-scattering theory \cite{NajiTHz2}, demonstrating good agreement. In addition to representing an extension of the forward-scatter theory by including the reflected waves, this theory helps to confirm the validity of the forward-scatter approximation \cite{NajiTHz2} for the analysis of overmoded structures.

Given the diffractive nature of the iris line and the shortness of a single-pass THz pulse (ignoring any secondary echoes), the diffraction loss is estimated by finding the ratio of the power scattered forward (down the line) to the power incident from the source \cite{NajiTHz2}. The part of the signal power that is scattered into the shadow regions between the screens is thus equivalent to what is ``lost" to diffraction. 

In the following numerical examples, we will take the example of a canonical structure with dimensions $a=55$~cm, $b=33.3$~cm, $r_0=2a$, and a total line length of $150$~m (equivalent to approximately 451 irises), which is a candidate iris-line structure for THz radiation transport at the SLAC National Accelerator Laboratory, Stanford University \cite{DESY,NajiTHz1,NajiTHz2,Zhang}. Since the diffraction loss is most severe at the lower frequency end of the operating range (3--15~THz), we study the 3~THz frequency ($\lambda=0.1$~mm) and we limit all our modal sums to 500 terms.   We will also observe the field output at consecutive cells in order to examine the effect of mode-launching purity for different sources and how the the excited fields gradually settle down toward the steady state of the structure. When the excitation is well matched to the structure's dominant eigenmode, mode purity is high and we observe the field settling to the steady state over a relatively short distance, with a desirable profile similar to the $J_0(2.4r/a)$ form and a uniformly linear polarization across the iris \cite{NajiTHz1, NajiTHz2}. If the source is mismatched, however, the signal, which is represented as a linear superposition of eigenmodes, will encourage mode competition within the structure, and will take longer to settle down to the desired mode at the output of the iris line.  This is demonstrated by the following four excitation scenarios.

\begin{figure}
	\includegraphics[width=0.94\columnwidth]{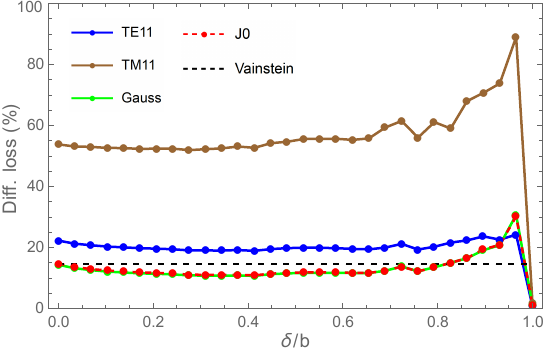}
	\caption{Behavior of diffraction loss as a function of screen thickness $\delta$, for four types of excitations: $J_0(2.4r/a)$, backed-off Gaussian, TE$_{11}$, and TM$_{11}$.}
	\label{fig: diff all}
\end{figure}

\subsection{Well-matched $J_0$ source} \label{J0 source}

Exciting the first iris of the line by a Bessel source with a dipolar transverse $E$ field profile of the form $J_0(2.4 r/a)$ represents an excitation that is well-matched to the ideal eigenmode (steady state) of the structure in the Vainstein limit. The latter limit occurs when the structure is oversized (overmoded) with a large Fresnel number, $N_f=a^2/(\lambda b)\gg 1$, and very thin screens, $\delta\rightarrow 0$, which leads to a transverse $E$ field \cite{DESY,Vainstein1} of the desirable form
\begin{equation}
J_{0}\left[\left(1-\varepsilon-i\varepsilon\right)\frac{2.4 r}{a}\right], \label{Vainstein J0}
\end{equation}
where $\varepsilon\approx0.164\sqrt{\lambda b}/a$ and is assumed small ($\varepsilon\ll 1$) in Vainstein's limit. This profile is akin to the $J_{0}(2.4r/a)$ profile, but with a small perturbation $\varepsilon$ that enters through its complex argument. The latter is a key feature of this geometry and represents the complex-impedance boundary condition (also named Vainstein's impedance condition \cite{DESY,NajiTHz1,NajiTHz2}), which is equivalent to the structure's periodic nature and enables the modes to mix and become hybrid in the axial region $r\leq a$.  Therefore, this excitation is expected to have the shortest transient regime, with the mode settling relatively quickly into the steady-state, as confirmed by Figure~\ref{fig: allPlots}(a). For very thin screens ($\delta\rightarrow 0$), the incurred diffraction loss at the end of the line for this excitation type is approximately 14\%, which is the lowest value among all four scenarios, and in agreement with Vainstein's limit (benchmark) \cite{DESY,Vainstein1,Vainstein2,NajiTHz1,NajiTHz2}.  

\subsection{Nearly matched backed-off Gaussian source} \label{Gauss source}

\begin{figure}
	\includegraphics[width=0.94\columnwidth]{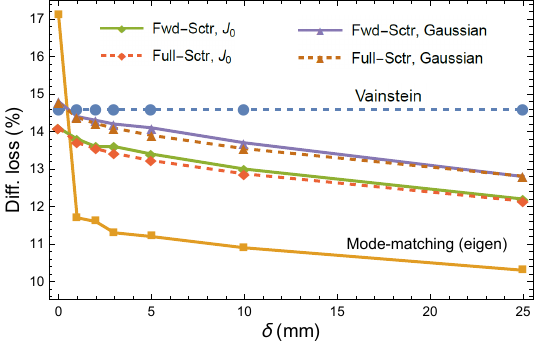}
	\caption{Comparing diffraction loss predictions from different theories, for the $J_0(2.4r/a)$ and backed-off Gaussian excitations. Good agreement is observed between the present full-scatter theory and the forward-scatter approximation \cite{NajiTHz2}.}
	\label{fig: comparison}
\end{figure}

A Gaussian dipole source is also considered, where we align the Gaussian beam's waist with the first iris, so as to have flat wavefronts at the iris's input plane. We control the Gaussian $e^{-r^2/w^2}$ profile by having the $1/e^2$-width backed off to $\approx0.65 a$. This gives a Gaussian profile of $\exp[-r^2/(0.65a)^2]$ that is better matched to the ideal $J_0(2.4 r/a)$ profile \cite{NajiTHz2}, for maximum power transfer. The transient regime for this case is shown in Figure~\ref{fig: allPlots}(b), showing evolution toward the eigenmode that is relatively quick, but not as quick as the $J_0(2.4r/a)$ excitation in Subsection~\ref{J0 source}. The overall diffraction loss for this case, using thin screens, is approximately 15\%, which is slightly higher than the $J_0(2.4r/a)$ excitation case, and in agreement with the forward-scatter theory developed in \cite{NajiTHz2}.

\subsection{TE$_{11}$ and TM$_{11}$ sources}
Exciting the iris line input with a TE$_{11}$ or TM$_{11}$ source, which are not well matched to the ideal eigenmode, will incur longer transient regimes, as observed in Figures~\ref{fig: allPlots}(c) and (d), where longer distances are needed for the transient regime to settle down and reach steady-state, especially for TM$_{11}$. For thin screens, the diffraction loss incurred at the end of the line is approximately 22\% for the TE$_{11}$ case, and 54\% for the TM$_{11}$ case. This is in agreement with the forward-scatter theory developed in \cite{NajiTHz2}.

\begin{figure}
	\includegraphics[width=0.94\columnwidth]{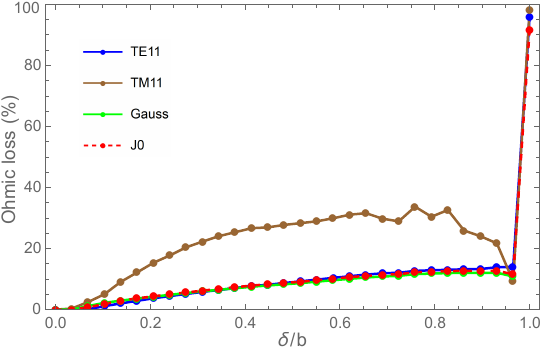}
	\caption{Behavior of ohmic loss (coherent modes) as a function of screen thickness $\delta$, for four types of excitations: $J_0(2.4r/a)$, backed-off Gaussian, TE$_{11}$, and TM$_{11}$. See text for further discussion.}
	\label{fig: ohmic all}
\end{figure}

\subsection{Effect of thicker screens on various modes}

By running the analysis for the same structure, $a=55$~cm, $b=33.3$~cm, $r_0=2a$, and a total line length of $150$~m (equivalent to 451 irises) at 3~THz, but with varying screen thickness $\delta$ relative to the period $b$, we can also observe how the diffraction loss and ohmic loss will scale as a function of thickness.  This is useful in practice, since thicker screens may be easier for mechanical design and construction. Figure~\ref{fig: diff all} provides a summary of the behavior of diffraction loss as a function of $\delta/b$, for the Bessel $J_0(2.4r/a)$, backed-off Gaussian, TE$_{11}$ and TM$_{11}$ excitations. We note how the Gaussian-source case curve tends to closely shadow that of the $J_0$-source (they are under each other in Figure~\ref{fig: diff all}). This can be readily explained by noticing the relatively quick settlement of both their transients in Figure~\ref{fig: allPlots}~(a) and (b).

A comparison between the present analysis and the different methods of analysis reported previously is given in Figure~\ref{fig: comparison}. This includes the Vainstein limit (at $\delta\rightarrow 0$) \cite{DESY,Vainstein1,Vainstein2,Vainstein3, NajiTHz1}, the mode-matching eigensolution analysis \cite{NajiTHz1}, and the forward-scatter field analysis \cite{NajiTHz2}. The results are shown for well matched sources, $J_0{2.4r/a}$ and the backed-off Gaussian. It is seen that the full-scatter and forward-scatter methods are in close agreement with each other in general, validating the use of forward-scatter approximation for such structure, and both agree well with Vainstein's limit for thin screens. The numerical implementation of the mode-matching eigensolution is seen to be in reasonable agreement as well, with approximation 2\% deviation from the two scattering theories.

\begin{figure}
	\includegraphics[width=0.94\columnwidth]{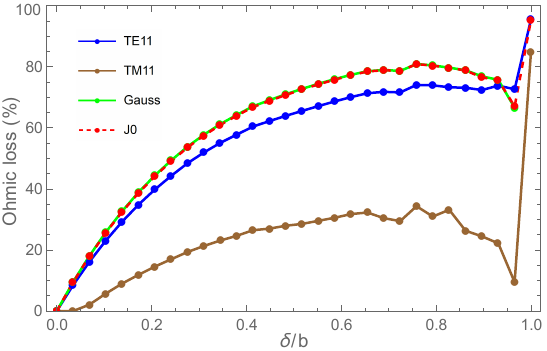}
	\caption{Behavior of ohmic loss (incoherent modes) as a function of screen thickness $\delta$, for four types of excitations: $J_0(2.4r/a)$, backed-off Gaussian, TE$_{11}$, and TM$_{11}$. See text for further discussion.}
	\label{fig: ohmic all2}
\end{figure}

Figure~\ref{fig: ohmic all} illustrates how ohmic loss in the signal's path scales with $\delta/b$, for the Bessel $J_0(2.4r/a)$, backed-off Gaussian, TE$_{11}$ and TM$_{11}$ modes, assuming coherent propagation. As in the case of diffraction curves in Figure~\ref{fig: diff all}, we note how the $J_0$ source curve and the Gaussian source curve are overlapping. As discussed in \cite{NajiTHz2}, this loss is attributed to the presence of azimuthal and longitudinal surfaces currents on the iris rims. Given the thin skin layer for good conductors at THz frequencies \cite{NajiTHz2}, the ohmic loss is found on each screen rim (of width $\delta$) using the traditional perturbation technique, which uses the tangential magnetic field when the surface is assumed to be a perfect conductor \cite{NajiTHz2,Collin,Collin2,jackson,zangwill,pozar}. Namely, the power loss per unit length at given point $z_0$ along the axial direction, $p'=\partial P/\partial z_{|_{z_0}}
$, is given by 
\begin{equation}
    p'=\frac{R_s}{2}\int\limits^{2\pi}_{0}\left[ |H_\phi(a,\phi,z_0)|^{2}+|H_z(a,\phi,z_0)|^{2} \right] a d\phi, \label{P0}
\end{equation} 
where $R_s=\sqrt{kZ_0/2\sigma}$ is the skin depth and $\sigma$ is the conductivity of the screen edge. The total magnetic fields, $H_\phi$ and $H_z$ in (\ref{P0}), are the sums that include all the modal terms of each field, $\sum_n H^\pm_{n\phi}$ and $\sum_n H^\pm_{nz}$. The rate $p'$ can be used in two ways to deduce the total ohmic loss incurred by the irises, depending on whether the modes that make up the THz pulse are traveling coherently or incoherently down the line. If the dispersion in modal group velocities is significant enough as to result in a temporal spread beyond the pulse's coherent interval, the modal contributions to the time-averaged ohmic power appear individually, each being representable by the usual attenuation coefficients ($\alpha_{\text{TE}_1n}$,  $\alpha_{\text{TM}_1n}$) \cite{NajiTHz2,jackson,Collin2}. This is then summed up across all modes to get the total loss, as detailed in equations (107)--(110) in reference \cite{NajiTHz2}.  However, if the summation was for coherently propagating modes, the terms $|\sum_n H^\pm_{n\phi}|^2$ and $|\sum_n H^\pm_{nz}|^2$ will include cross-terms among the modes, which can allow the different TE/TM modes to interfere together, partially canceling some of their individual surface-current contributions. For the current iris-line example, we expect coherent propagation for the modes of the THz pulse traveling paraxially through the oversized structure; see Figure~\ref{fig: ohmic all}. For reference, the loss calculated for the corresponding case of incoherent propagation is shown in Figure~\ref{fig: ohmic all2}, for the same iris-line example. Note that the estimation given by equations (107)--(110) in reference \cite{NajiTHz2} will therefore be in error (overerstimated) under coherent propagation, and one should instead sum the terms in $|\sum_n H^\pm_{n\phi}|^2$ and $|\sum_n H^\pm_{nz}|^2$ without assuming separable (incoherent) modal contributions. 

It is worth noting that the full-scatter analysis does not assume, for any reference plane of a given a step-in/-out discontinuity, at any given instant in time, that a backward-traveling wave exists on the right-hand side of the discontinuity prior to the arrival of the exciting THz pulse (incident from the left). As noted before, this is because we are mainly concerned with the first passage of the THz pulse through the system, with no considerable echoes (or other THz excitations) either before or after pulse under consideration. Indeed, using the principle of causality, the wave's paraxial incidence, the oversized dimensions of the iris line, and the fact that the pulse's duration $\tau$ is much shorter than the period $b$ and line length, it is clear that no backward-reflections from the right-side of the pulse will exist -- see \cite{NajiTHz2} for more details on this point. This is further confirmed by the fact that the full-scatter analysis has validated the forward-scatter analysis with negligible error, indicating very weak reflections, which could only interact at later times (after the main pulse has passed) with the rest of the structure (more reflections). This is a feature of paraxial propagation in the overmoded waveguide at the high-frequency limit, and is expected to hold even if the pulse duration was longer. Indeed, the reported behavioral agreement with the eigen-analysis \cite{NajiTHz1} indicates that this observation holds even at steady-state. If the structure is moved from the this regime into the low-frequency regime with comparable dimensions to the wavelength, then the traditional theory on multiple small reflections (see \cite{Collin,Collin2}, for example) may be utilized.

\section{Conclusions} \label{sec: conclusions}

The full-scatter vector field analysis for an overmoded iris-line structure, with paraxially incident THz radiation pulses, is presented. The spectral (modal) expansion of the full-scatter fields are derived and used to generate the scattering matrices of the waveguide sections. An advanced application of Lorentz's reciprocity theory, using a generalized guided-field configuration, is developed to reduce complexity of the mode-matching problem over nonuniform sections.  The method is implemented into numerical examples, with different source excitations, showing good agreement with previously established theories, including the Vainstein's limit (benchmark) and the forward-scatter approximation. The results are presented for four difference excitation types: $J_0(2.4r/a)$ Bessel, backed-off Gaussian, TE$_{11}$, and TM$_{11}$. Scaling trends for diffraction and ohmic losses are presented, as functions of $\delta/b$, for the four excitation types. This method provides a more accurate analytical framework for estimating the fields in overmoded iris-line structures, as it does not ignore the back-scattered field terms. The used technique is quite general and applies to a wide class of structures of similar configuration. It also validates the usage of the forward-scatter approximation for analyzing such structures at the high-frequency limit.

\begin{acknowledgments}
The authors would like to thank Karl Bane and Nour Elwagdi for several interesting discussions in relation to this research. This work was supported by the University of Santa Clara (GR103931-DENG1294Naji).

\end{acknowledgments}


\appendix


\section{\label{Appx1} Proof of Schelkunoff's equivalence theorem (\ref{Shelk-left field})--(\ref{Schek J source})}

Although Schelkunoff's theorem was not proved explicitly in \cite{schelkunoff}, one can easily prove it by looking carefully at the boundary conditions of the original and equivalent problems in Figure~\ref{Schelk}. 

\textit{Proof}: Since the transverse fields on the left- and right-hand sides of the aperture's plane must by continuous when the aperture in the diaphragm is open, the boundary condition $(\mathbf{E},\mathbf{H})_{z=0^-}=(\mathbf{E},\mathbf{H})_{z=0^+}$ must be satisfied. If the region $z>0$ is now occupied by a denser and denser medium, reaching the limit of perfect electric conductor (PEC) in that region, with the fields $(\mathbf{E},\mathbf{H})_{z>0}\rightarrow 0$, the fields in $z<0$ can be found and denoted by $(\mathbf{E},\mathbf{H})_{z<0}\rightarrow (\mathbf{E}_0,\mathbf{H}_0)$. The original field on the left side can thus be written as 
\begin{equation}
    (\mathbf{E},\mathbf{H})_{z<0}=(\mathbf{E}_0,\mathbf{H}_0)+(\mathbf{E}_c,\mathbf{H}_c) \label{field on left side},
\end{equation} where the field term $(\mathbf{E}_c,\mathbf{H}_c)$ represents a correction (difference) function. Note that the presence of the source in $z<0$ is still naturally represented by a singularity in the  $(\mathbf{E}_0,\mathbf{H}_0)$ equivalent field, and an analogous treatment can be carried out if the medium is filled with a perfect magnetic conductor. (Note, however, that since a knowledge of either $\mathbf{J}$ or $\mathbf{M}$ on the boundary is sufficient for uniqueness \cite{Stratton}, in this version of the theorem we consider only the current density $\mathbf{J}$.)  

However, with the PEC present, we now have a field discontinuity on the wall at $z=0$ (the diaphragm parts and PEC filling combined), which is accounted for by the appearance of a surface current  \begin{equation}
    \mathbf{j}_0=\hat{n}\times \left[ (\mathbf{H})_{z=0^+}-(\mathbf{H})_{z=0^-}\right]= -\hat{n}\times (\mathbf{H}_0)_{z=0^-}, \label{small j}
\end{equation} where the normal vector $\hat{n}$ is taken to be pointing to the right, as shown in Figure~\ref{Schelk}. The original boundary condition on the magnetic field can thus be re-written as follows (the associated electric field will follow from Maxwell's curl equations)
\begin{align}
    (\mathbf{H})_{z=0^-}&=(\mathbf{H})_{z=0^+}\nonumber\\
    (\mathbf{H}_0)_{z=0^-}+(\mathbf{H}_c)_{z=0^-}&=(\mathbf{H})_{z=0^+}\nonumber\\
    \hat{n}\times(\mathbf{H}_0)_{z=0^-}+\hat{n}\times(\mathbf{H}_c)_{z=0^-}&=\hat{n}\times(\mathbf{H})_{z=0^+}\nonumber\\
    \Rightarrow -\mathbf{j}_0=+\hat{n}\times\left[ (\mathbf{H})_{z=0^+} \right. & \left.-(\mathbf{H}_c)_{z=0^-}\right] \label{scattered}
\end{align}
We note, however, that the RHS of (\ref{scattered}) can be interpreted as nothing but a representation of field matching problem at a boundary at $z=0$, which happens to have the original field $(\mathbf{E,H})_{z>0}$ scattered to the right and the correction field $(\mathbf{E}_c,\mathbf{H}_c)_{z<0}$ scattered to the left, apparently due to an equivalent surface current density equal to $\mathbf{J}=-\mathbf{j}_0$ at the plane $z=0$ (with the original source and PEC now removed). The value of this current source can be obtained from the original problem when the PEC condition is present, since $\mathbf{J}=-\mathbf{j}_0=\hat{n}\times (\mathbf{H}_0)_{z=0^-}$. Thus, if we call the fields $(\mathbf{E,H})_{z>0}$ and $(\mathbf{E}_c,\mathbf{H}_c)_{z<0}$ scattered by $\mathbf{J}$ as $(\tilde{\mathbf{E}}^+,\tilde{\mathbf{H}}^+)$ and $(\tilde{\mathbf{E}}^-,\tilde{\mathbf{H}}^-)$, respectively, and substitute in (\ref{field on left side}), we see that the original fields can be given by
\begin{align}
    (\mathbf{E},\mathbf{H})_{z<0} &= (\mathbf{E}_0,\mathbf{H}_0) + (\tilde{\mathbf{E}}^-,\tilde{\mathbf{H}}^-) \nonumber\\
    (\mathbf{E},\mathbf{H})_{z>0} &= (\tilde{\mathbf{E}}^+,\tilde{\mathbf{H}}^+),\nonumber
\end{align}
which is the main result of the theorem, as seen in equations (\ref{Shelk-left field}) and (\ref{Shelk-right field}), and completes the proof.

\hfill

\bibliography{myrefs}

\end{document}